\pdfoutput=1

\documentclass[8.5pt,twoside,twocolumn]{article}
\usepackage[super,sort&compress,comma]{natbib} 
\usepackage{mhchem}
\usepackage{times,mathptmx}
\usepackage{natbib}
\usepackage{color}

\usepackage{sectsty}
\usepackage{balance} 
\usepackage{widetext}
\usepackage{IEEEtrantools}
\usepackage{url}

\usepackage[pdftex]{graphicx}  
\usepackage{epstopdf} 
\usepackage{pslatex} 
\usepackage{lastpage}
\usepackage[format=plain,justification=raggedright,singlelinecheck=false,font=small,labelfont=bf,labelsep=space]{caption} 
\usepackage{fancyhdr}
\usepackage{amssymb}

\begin{document}

\thispagestyle{plain}
\fancypagestyle{plain}{
\fancyhead[L]{\includegraphics[height=8pt]{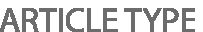}}
\fancyhead[C]{\hspace{-1cm}\includegraphics[height=20pt]{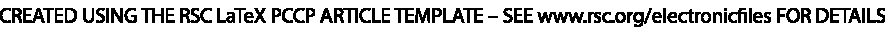}}
\fancyhead[R]{\includegraphics[height=10pt]{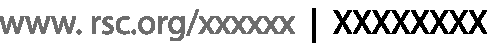}\vspace{-0.2cm}}
\renewcommand{\headrulewidth}{1pt}}
\renewcommand{\thefootnote}{\fnsymbol{footnote}}
\renewcommand\footnoterule{\vspace*{1pt}%
\hrule width 3.4in height 0.4pt \vspace*{5pt}} 
\setcounter{secnumdepth}{5}

\makeatletter 
\def\subsubsection{\@startsection{subsubsection}{3}{10pt}{-1.25ex plus -1ex minus -.1ex}{0ex plus 0ex}{\normalsize\bf}} 
\def\paragraph{\@startsection{paragraph}{4}{10pt}{-1.25ex plus -1ex minus -.1ex}{0ex plus 0ex}{\normalsize\textit}} 
\renewcommand\@biblabel[1]{#1}            
\renewcommand\@makefntext[1]%
{\noindent\makebox[0pt][r]{\@thefnmark\,}#1}
\makeatother 
\renewcommand{\figurename}{\small{Fig.}~}
\newcommand{\RP}[1]{{#1}}

\def\topfraction{.9}
\def\floatpagefraction{.8}

\sectionfont{\large}
\subsectionfont{\normalsize} 

\fancyfoot{}
\fancyfoot[LO,RE]{\vspace{-7pt}\includegraphics[height=9pt]{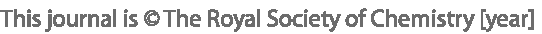}}
\fancyfoot[CO]{\vspace{-7.2pt}\hspace{12.2cm}\includegraphics{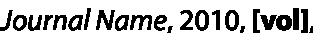}}
\fancyfoot[CE]{\vspace{-7pt}\hspace{-13.5cm}\includegraphics{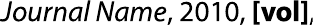}}
\fancyfoot[RO]{\footnotesize{\sffamily{1--\pageref{LastPage} ~\textbar  \hspace{2pt}\thepage}}}
\fancyfoot[LE]{\footnotesize{\sffamily{\thepage~\textbar\hspace{3.45cm} 1--\pageref{LastPage}}}}
\fancyhead{}
\renewcommand{\headrulewidth}{1pt} 
\renewcommand{\footrulewidth}{1pt}
\setlength{\arrayrulewidth}{1pt}
\setlength{\columnsep}{6.5mm}
\setlength\bibsep{1pt}

\twocolumn[
  \begin{@twocolumnfalse}
\noindent\LARGE{\textbf{Chirality-Dependent Properties of Carbon Nanotubes: \\ Electronic Structure, Optical Dispersion Properties, Hamaker \\
Coefficients and van der Waals -- London dispersion interactions$^\dag$}}
\vspace{0.6cm}

\noindent\large{\textbf{Rick F. Rajter $\dagger\dagger$, Roger H. French $\S$, W.Y. Ching$\S\dagger$, Rudolf Podgornik$\ddagger$$\star$ and V. Adrian Parsegian$\ddagger$}}\vspace{0.5cm}

\noindent\textit{\small{\textbf{Received Xth XXXXXXXXXX 20XX, Accepted Xth XXXXXXXXX 20XX\newline
First published on the web Xth XXXXXXXXXX 200X}}}

\noindent \textbf{\small{DOI: 10.1039/b000000x}}
\vspace{0.6cm}

\noindent \normalsize{Optical dispersion spectra at energies up to 30 eV play a vital role in understanding the chirality-dependent van der Waals – London dispersion interactions of single wall carbon nanotubes (SWCNTs). We use {\RP one-electron theory based calculations} to obtain the band structures and the frequency dependent dielectric response function from 0-30 eV for 64 SWCNTs {\RP differing in radius, electronic structure classification, and geometry}. The resulting optical dispersion properties can be categorized over three distinct energy intervals (M, $\pi$, and $\sigma$, respectively representing 0-0.1, 0.1-5, and 5-30 eV regions) and over radii above or below the zone-folding limit of 0.7 nm.  While $\pi$ peaks vary systematically with radius for a given electronic structure type, $\sigma$ peaks are independent of tube radius above the zone folding limit and depend entirely on SWCNT geometry. We also observe the so-called “metal paradox”, where a SWCNT has a metallic band structure and continuous density of states through the Fermi level but still behaves optically like a material with a large optical band gap between M and $\pi$ regions. This paradox appears to be unique to armchair and large diameter zigzag nanotubes. Based on these {\RP calculated one-electron dielectric response functions} we compute and review Van der Waals - London dispersion spectra, full spectral Hamaker coefficients, and van der Waals - London dispersion interaction energies {\RP for all calculated frequency dependent dielectric response functions}. Our results are categorized using a new optical dielectric function classification scheme that groups the nanotubes according to observable trends and notable features (e.g. the metal paradox ) in the 0-30 eV part of the optical dispersion spectra. While the trends in these spectra begin to break down at the zone folding diameter limit, the trends in the related van der Waals - London dispersion spectra tend to remain stable all the way down to the smallest single wall carbon nanotubes in a given class.}
\vspace{0.5cm}
 \end{@twocolumnfalse}
  ]

\section{Introduction}
\label{sec:into}

\footnotetext{\dag~Electronic Supplementary Information (ESI) available: [details of any supplementary information available should be included here]. See DOI: 10.1039/b000000x/}

\footnotetext{\textit{$\dagger\dagger$ Department of Materials Science and Engineering, Massachusetts Institute of Technology, Cambridge, Massachusetts 02139-4307, USA}}
\footnotetext{\textit{$\dagger$  Department of Materials Science and Engineering, Case School of Engineering, Case Western Reserve University, Cleveland Ohio, 44106-7204, USA}}
\footnotetext{\textit{$\S\dagger$Department of Physics, University of Missouri-Kansas City, Kansas City, Missouri, 64110, USA }}
\footnotetext{\textit{$\star$Department of physics, Faculty of Mathematics and Physics, University of Ljubljana,1000 Ljubljana, Slovenia and Department of Theoretical Physics, J. Stefan Institute, 1000 Ljubljana, Slovenia}}
\footnotetext{\textit{$\ddagger$Department of physics, University of Massachusetts, Amherst MA 01003, USA  (corresponding author: parsegian@physics.umass.edu) }}

{\RP Since their discovery in 1993 by the Iijima and the Bethune groups} \cite{Iijima1993,Bethune1993}, single wall carbon nanotubes (SWCNTs) have received enthusiastic attention.  Their intriguing mechanical and electrical properties make them ideal for a wide range of applications, from metal oxide semiconductor field effect transistors (MOSFETs) to novel drug delivery systems \cite{[2], [3], [4], [5], [6], [7]} and  have inspired strong scientific and industrial interest \cite{Iijima1993,[15],XYKong,CABessel,JPKim,Srivastava,Hayamizu,JDai}. The area that has arguably received the most attention and focus is their electronic structure and characterization \cite{Popov_CNT_overview,[14],Mintmire_metal_semi,[39]}. The reason is that SWCNTs are a unique class of materials in which small changes in their atomic structure and geometry characteristics (quantified by the chirality vector {\RP (n,m)} or equivalently by {\RP [n,m, type]}, when we want to stress the type of the nanotube \footnote{{\RP In the context of the dielectric response functions of CNTs, we denote the SWCNT with a mnemonic [n,m, type] rather then the standard but equivalent (n,m), when we want to invoke its type explicitly in order to help the reader organize the voluminous information and in order to be consistent with our previous publications. Of course the two designations, together with the simple (n-m)/3 = integer rule to identify the metallic and semiconducting nanotubes \cite{[15]}, are equivalent.}}) determines whether they behave as a dielectric, a metallic conductor or a semiconductor. They are also very narrow (typically $\leq 1$nm wide) and inexpensive to produce, making them ideally suited for many nano-electronic applications while widespread commercial and industrial use has yet to be achieved. {\RP While there has been some recent progress in the purification of SWCNT  from residual amorphous carbon or bundled SWCNTs, that sometimes also leads to diameter-selective separation of SWCNTs \cite{Kataura}, it is in general still difficult to 
place them into predetermined arrangements \cite{[8], [9], [10], [11], [12]}. Solving the 
assembly problem, which has been addressed on various levels beginning shortly after the discovery of CNTs \cite{Assembly}, presents an important next step based on the detailed molecular structure of various SWCNTs that determines their optical properties and through them their interactions: a subject of this review. }

Understanding the optical dispersion properties of SWCNT is crucial to their role in several long-range interactions, especially  the van der Waals - London dispersion (vdW-Ld) interaction. Each SWCNT is uniquely defined by its chirality vector {\RP (n,m)}, which denotes the circumferential wrapping direction along a graphene sheet containing $\rm sp^2$ hybridized carbon bonds \cite{[13], [14], [15]}.  Thus each {\RP (n,m)} combination denotes a different geometrical construction that is responsible for the wide variety of electronic structure, phonon, and Raman behavior \cite{[16], [17], [18]}.  The same diversity that makes them appealing also for performance leads to selectivity and specificity in sorting and placement. This is seen in AC/DC experiments \cite{[19]}, dielectrophoresis \cite{[20]}, and anion IEC \cite{[10], [11], [12]}. Each chirality vector gives SWCNTs a unique set of properties (e.g., band gap, van Hove singularities (vHs), etc.), but their common composition and chemical bonding coupled with symmetry considerations makes it possible to categorize them by {\RP their particular features \cite{[13], [18]}.
	
	
	A comprehensive study of CNT} interactions can provide the fundamental conceptual framework required for proper understanding of experiments on their separation, placement and assembly. Among these interactions, the universal and longest range, the van der Waals-London dispersion (vdW-Ld) interaction, merits most detailed scrutiny. First, because these interactions are long range and ever-present in nature, always contributing to and influencing the overall interaction between charge-neutral bodies \cite{French-RMP}, second, previous results \cite{[46],[48],Toni} show that the magnitude of this interaction for various CNTs can vary by 20\% even in simple systems, making it a convenient candidate to control experimentally, and third, SWCNTs have well known trends in their electronic structures as a function of chirality \cite{[39]}, which could then act as the source for chirality-specific vdW-Ld interactions. 

	During the last few years it has become possible to compute quantitative direction-dependent optical dispersion properties, such as the frequency dependent complex dielectric function $\varepsilon(\omega) = \varepsilon'(\omega) + \imath \varepsilon''(\omega)$, for a diverse set of isolated SWCNTs, for which spectroscopic techniques even as powerful as electron energy loss spectroscopy (EELS) \cite{[21], [22]}, which requires bulk sample of identical SWCNTs in a bundle \cite{[23]}, cannot be applied directly. {\RP The use of one-electron theory based calculations allows us} to circumvent these shortcomings but still use available experimental results to compare and confirm the integrity of our spectral data \cite{[24], [25], [26]}. We were able to resolve the axial and radial direction of SWCNTs to obtain the optical dispersion properties out to photon energies of at least 30 eV.  Also, we were not limited to the SWCNTs that are easily produced by existing preparation methods \cite{[27], [28]}, which allowed us to study a larger set of data than would otherwise be possible.  

	Early attempts to obtain this information via {\sl ab initio} codes produced unreliable results above 20 eV and were limited to a single nanotube \cite{[25]}.  Some computational studies were able to investigate a large quantity of chiralities, but they were confined to dispersion properties below 6 eV  because they relied on the tight binding approximation, which artificially distorts the bands \cite{[29]}.  New {\sl ab initio} results are emerging that have accurate dispersion properties along the entire interval, but continue to be confined to a select number of chiralities \cite{[26], [24]}.  Here we calculate and review the optical dispersion properties, represented as the imaginary part of the dielectric response function, $\varepsilon''(\omega)$, up to 30 eV for 64 SWCNTs that have varying diameter, electronic structure classification, and geometry. We follow trends and variations to give insight for system design. Later, we use these dielectric response functions to analyze consequences of different spectral properties in vdW-Ld interaction energies.

Trends and deviations from these trends in SWCNT optical dispersion spectra are categorized into a new SWCNT classification scheme,  based on a combination of electronic structure, geometry and curvature. The links between chirality, cuttings lines, band diagrams, density of states and optical dispersion properties are extended to the magnitude of vdW-Ld interaction as quantified by the Hamaker coefficients for chirality-dependent vdW-Ld interactions. We highlight the key remaining features linking the {\RP dielectric function $\varepsilon''(\omega)$, with the consequent} vdW-Ld interaction energy. 


\section{Electronic structure and optical dispersion spectra}
\label{sec:electronic_structure}

\subsection{Method}

The method for calculation of the electronic structure and optical properties is the first principles orthogonalized linear combination of atomic orbital (OLCAO) method which is described in detail in Ref. \cite{WaiYimBook}. The OLCAO method is applied to the calculation of the electronic structure and optical dispersion spectra of 64 SWCNTs.  {\RP It yields the imaginary part of the frequency-dependent dielectric response function, $\varepsilon''(\omega)$, 
calculated within the random phase approximation (RPA) of the one-electron theory \cite{[38]}. 
The optical transitions from the valence band (VB) to the conduction band (CB) are calculated from the {\sl ab initio} wave functions using the extended basis (EB). 
The {\sl ab initio} nature of the wave function ensures the strict adherence of the selection rules via the momentum matrix elements. If an optical transition is symmetry forbidden, it automatically has a zero momentum matrix and no transition will occur even if such a transition appears to be available via the density of states (DOS). A key example that we will show later is when a SWCNTs DOS has a continuous series of states across the Fermi level and yet behaves like a semiconductor with no transitions below 1 eV. All the relevant details of the calculation have been described in \cite{[46],[47],[48]}.}

{\RP Though not taken into account by the one-electron method, we are fully aware of the importance of the excitonic effects \cite{EXCITON,Dresselhous2007}  in the low dimensional systems such as graphene and CNTs \cite{Exciton1,Exciton2,Exciton3,Exciton4,Exciton5,Ando1997,Kane2003,Perebeinos2004,Pedersen2004,Chang2004,Maultzsch2005,Wang2005,Jiang2007,Exciton7,Exciton8}. They tend to introduce additional peaks in the energy range below 5 eV or slightly shift the position of some other peaks. From a pure theoretical point of view, the effect of excitons has been intensively studied and can be approached from many different angles including two-photon excitation, excitons of delocalized higher levels etc..  An important question in this respect is to what extent the excitonic peaks can modify the optical absorption obtained from the one-electron theory?  In some cases the traditional one-electron calculation of optical properties gives dielectric constants or the refractive indices in good agreement with experiments on many insulating materials.  However, though the excitonic effects in the case of SWCNT optical spectra are important and eminently measurable \cite{[16],[17]}, it is nevertheless not clear how the changes in {\sl optical spectra} due to excitonic effects will affect the overall {\sl Hamaker coefficients} and the trends in the vdW-Ld interaction between different classes of SWCNTs. Since the vdW-Ld interaction is a functional (or more precisely a discrete sum over the Matsubara frequencies) of the frequency-dependent dielectric response, it is reasonable to expect that such effects will be small and that the strength of the interaction will depend more on the global properties of the dielectric response delimited by the sum rule then on local specifics.  (see later results in section 6.3).  

The above conjecture is supported by the recent work of Hobbie and co-workers on empirical evaluation of attractive vdW potential in SWCNTs \cite{Exciton6}. Even though they have also recognized the presence of excitonic effect in CNTs, the authors used a much simpler empirical approach by fitting the collective peaks to a superposition of individual Lorentzians. They have concluded that for the semiconducting nanotubes, neglecting the three optical resonances reduces the Hamaker coefficient by roughly 5\%, while neglecting the Drude term decreases the Hamaker coefficient by roughly 2\%. For the metallic SWCNTs, neglect of either of these terms reduces the Hamaker coefficient by roughly 3\%. In this view, excitonic effects certainly have a small but measurable effect and a larger question is now how they influence the plasmons. From the {\sl ab initio} perspective, a first-principles approach that correctly accounts for electron-hole interactions in quasi-1D systems would thus provide a more accurate computational foundation, but only to the extent that it also accurately describes the two plasmons. On the other hand, practical applications to real problems using optical spectra obtained from one-electron calculation is an attractive alternative as indicated in another recent paper on adhesion in silicon wafers \cite{Loskill2012}. }

	The fundamental theory of the calculation method within the framework of the DFT is sound. Although many-body interactions are not considered and there may be concerns about the underestimation of the band gap, they are less important in this study since our main focus is on the {\RP vdW-Ld spectra} of a large number of CNTs up to high frequency and their systematic correlations to the structure. In contrast, in the tight binding approximation (TBA) method where no accurate wave functions are involved, the band gaps are typically adjusted in order to best describe and represent the states very near the band gap.  States several eV beyond the band gap are ignored or poorly represented \cite{[29]}. In addition, TBA models would not be able to predict novel departures from the main SWCNT structural trends.  There are several examples of this, such as 1) the [5,0,s] actually is a metallic conductor instead of being semiconductor-like \cite{[29]}, and 2) the [6,0,m] has a large degree of  hybridization of graphene’s $\rm sp^2$ and $\pi$ bonds into $\rm sp^3$ bonds and thus a band diagram unlike any other armchair SWCNT \cite{[18]}.  {\sl Ab initio} codes capture these major departures from the trends without requiring {\sl ipso facto} arbitrary adjustments.  On the other hand, inclusion of excitonic corrections based on many-body theory is not feasible in the OLCAO method which uses local atomic orbitals for basis expansion rather than plane waves. It is interesting to note that empirical TBA can also be refined to investigate the excitonic effect in SWCNT as has been demonstrated by the work of Jiang et al. \cite{Jiang2007}.

 The OLCAO method is an all-electron method with an economical use of the basis expansion. The full secular equation is diagonalized to obtain all electron states including high energy CB states. If an iterative procedure is used for the electron states, only a small number of CB state wave functions are usually obtained. To obtain higher energy CB states in an iterative approach for systems with a large number of atoms in the unit cell could be computationally prohibitive.   The dipole matrix elements of transitions between VB and CB are accurately calculated from the {\sl ab initio} wave functions and explicitly included in the optical dispersion spectra calculation, which automatically impose the selection rules for the transition. 

The OLCAO method is versatile and can be applied to almost any material whether it is a metal or an insulator, a liquid or an amorphous solid, an inorganic crystal or a biomolecule, having an open structure or a compact one, or whether it contains light elements (such as H or Li) or heavy atoms (such as rare earth elements).  The method avoids the use of atomic radii in the calculation, which could be problematic for complex materials with different local bonding. Finally, the use of local atomic orbitals facilitates the interpretation of the physical properties where $\pi$ or $\sigma$ states play a crucial role in the optical spectra of SWCNTs. The efficient evaluation of multicenter integrals in analytic forms makes the method highly efficient and eventually applicable to large complex systems such as DNA molecules in a fluid medium.  {\RP For example, application to the calculation of large anisotropic optical absorption in herapathite crystal ($\rm (C_{20}H_{24}N_{2O2}H_2)_4 \times C_2H_{4O} 2 \times 3SO_4 \times 2 I_3 \times 6 H_2O$, a complex multi-component organic crystal with 988 atoms in the unit cell yield excellent agreement with the measured data \cite{[63]}. Application of the OLCAO method to the spectroscopic properties of a large model of super-cooled water is equally successful \cite{Liang2011}}. 

\subsection{Case Study: SWCNT: [7,5,s] vs. [8,2,m]}

Each SWCNT  {\RP chirality (n,m)} defines a unique geometrical construction that results in a distinct set of electronic and optical dispersion properties. Because of their similarity to graphene in composition and structure, there are some features common among all SWCNTs, however, there are also two fundamental differences: circumferential periodicity and curvature. Periodicity along the circumference allows only discrete lines of states out of the continuum available for graphene, while curvature placed upon a flat graphene sheet re-hybridizes the $\rm sp^2$ lattice to include $\rm sp^3$ like states \cite{[39]}.

For an infinite diameter SWCNT, the number of available states is a continuum equivalent to nearly flat $\rm sp^2$ sheets of graphene. The only differentiating feature among various infinite diameter tubes would be the bonding direction along the axial direction. For a large, but finite diameter SWCNT, the effects of periodicity would cause a set of discrete allowable states in the axial direction. However, because the carbon bonds would still be relatively flat (i.e., still nearly perfect $\rm sp^2$ bonds), it is possible to derive a SWCNT’s band structure by selecting the allowable states from the band diagram of graphene. Such an approximation is called the “zone-folding” approximation; it is valid only for SWCNTs of diameter greater than 1 nm.  Below this threshold, the curvature of the SWCNT wall becomes so pronounced that the bonds gain a significant $\rm sp^3$ characteristic.  For very small tubes, using a zone folding procedure can be problematic \cite{[18], [29], [39], [40]}. 

\begin{figure*}[!t]
\begin{center}\includegraphics[width=12cm]{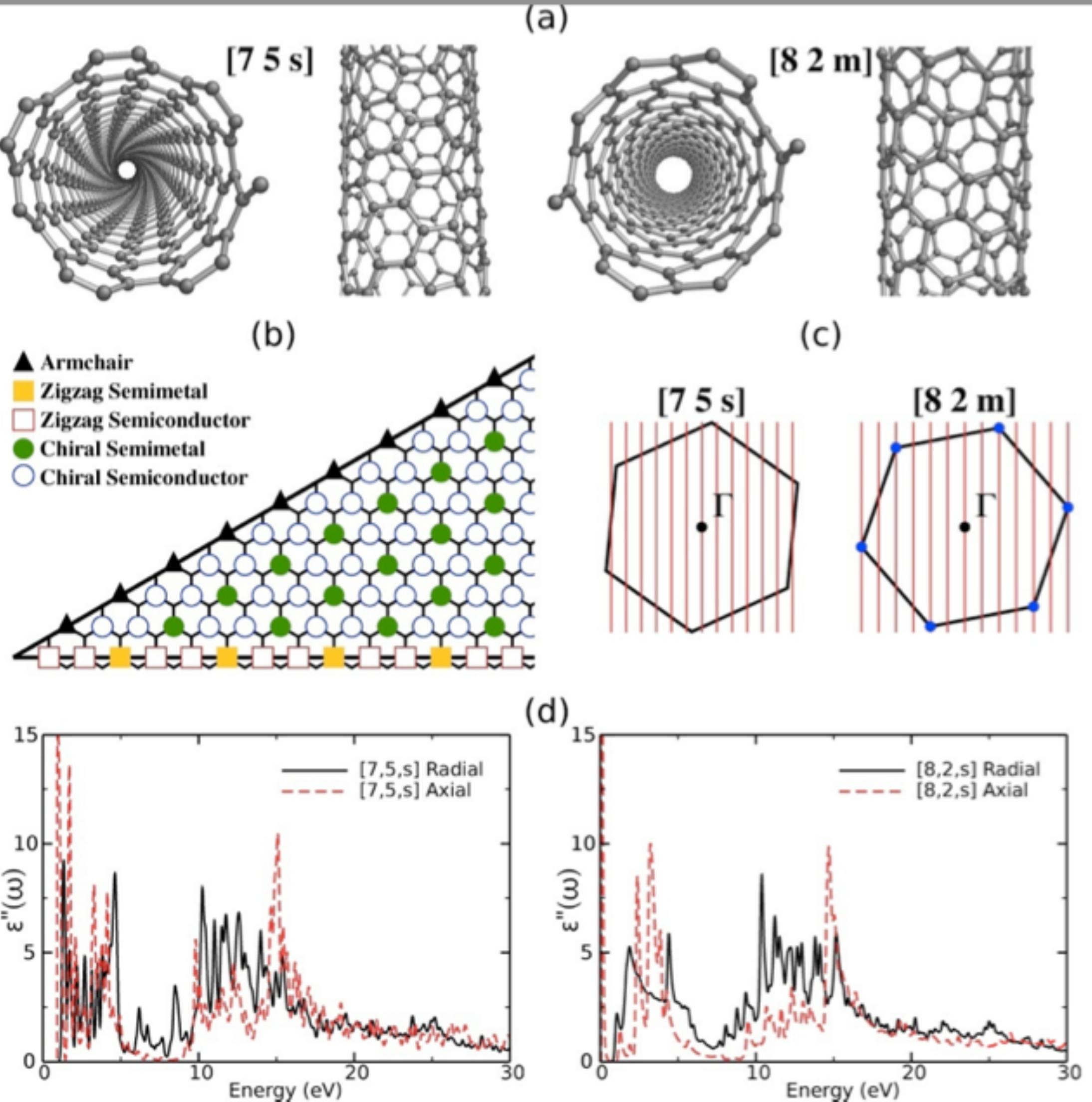}\end{center}
\caption{(color on line)  Layers of abstraction of SWCNT properties ranging from [n,m] to the optical dispersion spectrum $\varepsilon"(\omega)$ for the [7,5,s] semiconductor and [8,2,m] metal. (a) The geometrical construction. (b)	A chirality map showing all 5 types of SWCNT. (c) {\RP The cutting lines across the Brillouin zones for two different types of SWCNT, [7,5, s]  and [8, 2, m], within the dispersion classification. The dot represents the K point in the Brillouin zone.} (d)	The optical dispersion spectrum $\varepsilon"(\omega)$ in both the radial and axial direction.  Frequency in eV. }
\label{[figure4]}
\end{figure*}

Each SWCNT chirality represents a unique geometrical construction of a 1D tube of $\rm sp^2$ carbon bonds created from a rolled up sheet of graphene. What makes SWCNTs such a highly studied set of materials is the number of material properties that significantly change as a function of chirality. Perhaps the best well known is metallic versus semiconducting behavior, depending whether or not $\rm (n-m)/3$ is an integer. Variations in properties propagate through several features, for example: linkages between chirality, cuttings lines, band diagrams, density of states, optical dispersion properties, Hamaker coefficients, extending to chirality-dependent vdW-Ld interactions.  Figure \ref{[figure4]} shows these differences for the metallic [8,2,m] and the semiconducting [7,5,s].
	
	While some of these relationships (e.g., chirality to electronic band diagram) have been extensively studied and some predictive rules are widely used, much information at the optical level cannot be easily predicted {\sl a priori} or explained simply by knowing {\RP (n,m)}. Perhaps the most notable example is between the electronic structure and optical levels, which exhibit many non-linear relationships and other features that offer somewhat surprising results. The need is to determine the optical dispersion properties for most chiralities rather than approximate/derive them from a SWCNT of seemingly similar characteristics.  

	The introduction of {\sl ab initio} $ \varepsilon''(\omega)$ optical dispersion properties enables such an analysis, which would be very difficult experimentally. Fortunately, the experimental data that do exist (notably the EELS data by Stephan et al. \cite{[21]}) compare favorably to our results. Although they specify tubes by diameter and not chirality, they do note major peaks that appear consistently across tubes of all diameters. Most notably, their results show three major peak areas: band gap, transitions at ~4 eV, and transitions at ~15 eV.  There are also low flat areas between 5-12 eV and beyond 20 eV.  These results are in qualitative agreement with our [7,5,s] and [8,2,m] {\sl ab initio} optical dispersion properties.  The only major difference appears to be in the balance of the $\varepsilon''(\omega)$  peak heights and total areas within the $\pi$ and $\sigma $ regions. However, these differences may resolve when experimental data can be more readily divided into the radial and axial components, or perhaps a finer experimental energy resolution would help.  

\subsection{SWCNT classification schemes}

	Table 1 lists the 64 SWCNTs studied here, with their figures of merit (n, m, radius, angle, geometry, and number of atoms per lattice repeat), and the three classification schemes (zone folding, Lambin \cite{[39]}, and Saito \cite{[15]}).  Our proposed “dispersion” classification (not in Table 1) is essentially a combination of the Lambin and geometrical descriptors.  Although this technically leads to nine possible combinations, only five of them are allowed for smaller diameter SWCNTs. For instance, an armchair SWCNT’s electronic structure is never semiconducting unless there are added defects, functional groups, or other influential external source such as a strong external field \cite{[50], [51], [52], [53]}.  For the larger-diameter limit, the divisions may become redundant as the semimetal gap becomes so small (i.e., less than 0.01 eV) that it essentially blurs the line of the metal/semimetal division.  
	


\begin{figure}
\includegraphics[width=9cm]{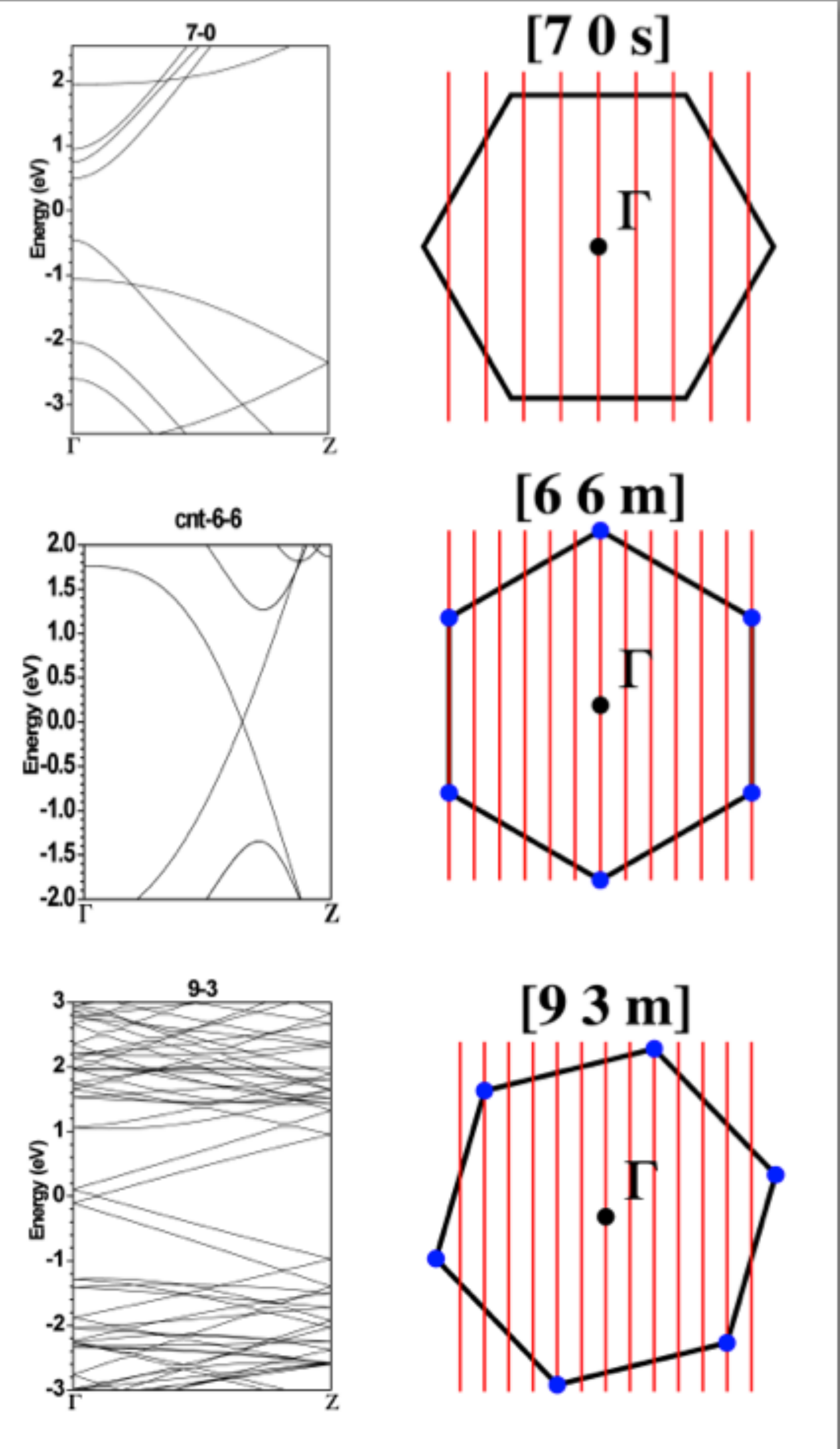}
\caption{(color on line) A comparison of the band diagrams and cutting lines for three SWCNTs 
representing the Lambin classification system.  Note the subtle difference between the [6,6,m] and the [9,3,m].  The [6,6,m] has no band gap whereas the [9,3,m] has CB and VB just miss touching. Thus the [9,3,m] is actually a “small band gap semiconductor” .}
\label{[figure6]}
\end{figure}

{\RP Figure \ref{[figure4]} shows the cutting lines for the SWCNTs representing an example of two types of the dispersion classification,  [7,5, s]  and [8, 2, m].  It is the relationship of cutting line angle, location, and density that determines if and when the cutting lines cross/graze the K points at the edge of the BZ and result in a metal/semimetal electronic structure.  Nanotubes of very different cutting line density and angle (i.e., the [6,5,s] and the [9,1,s]) can have different cutting line angles and yet similar band gaps. Conversely, tubes that appear to have very similar cutting line angles and density (i.e., the [9,0,m] versus the [10,0,s]) fall into different classifications altogether.  Such is the delicate nature of the electronic structure’s dependence on {\RP (n,m)}. }The cutting line representation in the BZ has served as the basis for every major SWCNT classification thus far (i.e., simple, Lambin, and Saito). In the zone-folding classification, the effects of curvature on the carbon $\rm sp^2$ bonds are ignored and the only thing that differentiates the basic metallic and semiconducting classifications is whether a K point is crossed.  This results in the simple (n-m)/3 = integer rule \cite{[15]}. The Lambin system acknowledges the changes in $\rm sp^2$ bond angles and lengths as a result of the cylindrical nature of SWCNTs \cite{[39], [18]}.  This curvature causes a slight distortion in the structure and the resulting BZ contains cutting lines that barely miss or “graze” the K-points and don’t directly crossing them.  This opens up the tiniest of band gaps (ranging from 0.02 to 0.05 eV) for tubes that the zone-folding classification would predict as perfect metals with no band gaps. Figure \ref{[figure6]} illustrates the cutting line diagrams and resulting electronic band structures of these three types.  The [9,3,m] has a barely visible 0.02 eV band gap. For comparison, the Lambin model predicts a gap of approximately 0.07 eV.

Finally, the Saito method \cite{[15]} returns to the zone-folding scheme (i.e., ignores the effects of the structural relaxation), but it considers additional symmetry effects that can further differentiate tubes into numerous groupings \cite{[15]}.  For example, the semiconductors can be subdivided into two classes based on the relationships of the first cutting line’s position relative to the K point.  
Neither the proposed dispersion classification nor the available classifications can adequately predict the dispersion properties of the smallest SWCNTs.  At this limit, there is a strong reintroduction of $\rm sp^3$-like bonding, which can engender new bands within the electronic structure as well as significantly shift other features.  For this reason, [5,0,s] is metallic despite no classification predicting it \cite{[29]}.  It appears, however, that there are very few of these exceptions.  Therefore, one should always seek additional experimental or {\sl ab initio} confirmation when describing the dispersion properties of very small SWCNTs.

\subsection{Influence of structural parameters on optical transitions}

\begin{figure*}[t!]
\begin{center}
\includegraphics[width=14cm]{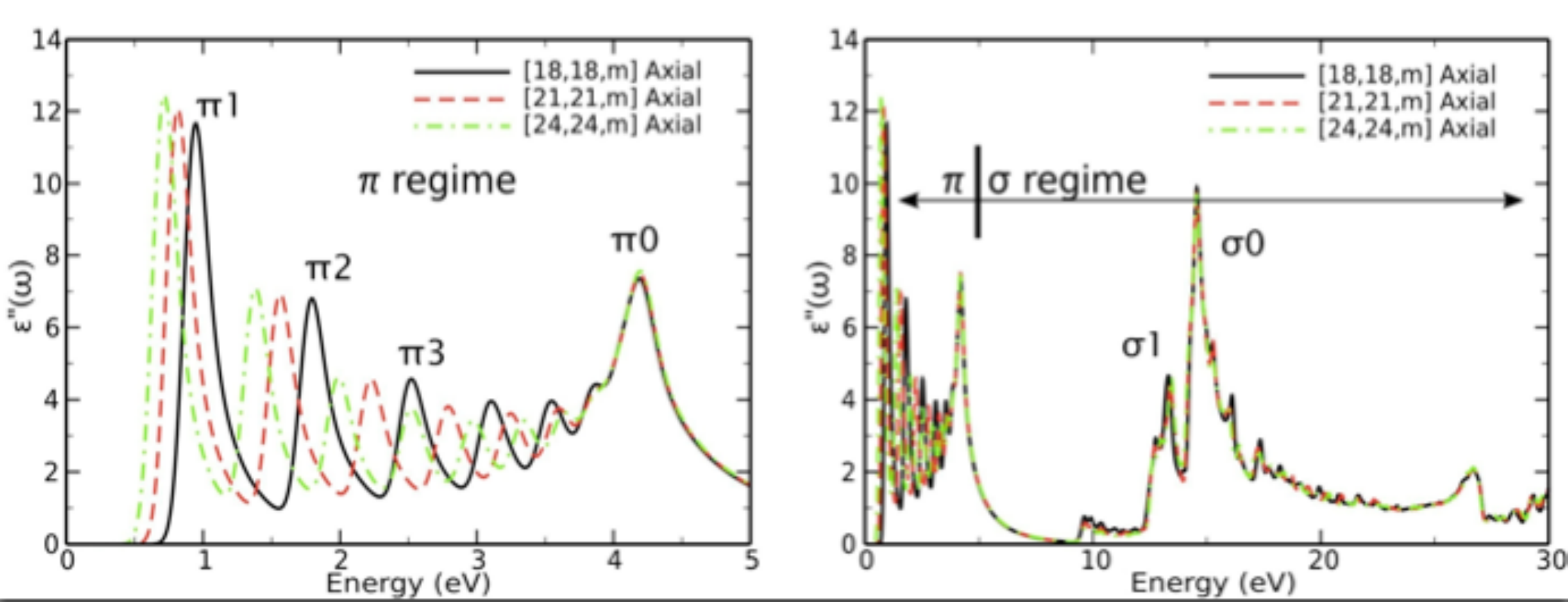} \end{center}
~~~~~~~~~~~~~~~~~~~~~~~~~~~~~~~~~~~~~~~~~~~~~~~~~~~~~~~~~~~~~~~(a)~~~~~~~~~~~~~~~~~~~~~~~~~~~~~~~~~~~~~~~~~~~~~~~~~~~~~~~~~~~~~~~~~~~~~~~~~~~~~(b)
\caption{(color on line)  {\RP The axial component of the optical dispersion spectrum $\varepsilon"(\omega)$. 
(a)	$\varepsilon"(\omega)$ in the 0-5 eV region. (b)	$\varepsilon"(\omega)$ in the 0-30 eV  region, both for the case of [18,18,m], [21,21,m] and [24,24,m] SWCNTs. }Note that as the cutting line density increases for larger tubes, there is a systematic shift of the first $\varepsilon"(\omega)$  $\pi_1$ peak towards 0eV. However, the optical dispersion spectrum $\varepsilon"(\omega)$ above 5 eV appear to be largely invariant with respect to changes in diameter for large tubes. Frequency in eV. }
\label{[figure7]} 
\end{figure*}

Despite the shortcomings in each of the given classifications, they are powerful tools that can help to map out and understand global trends. The first trend is that the packing density of cutting lines varies systematically as a function of radius, creating a diameter-dependent band gap and trends in {\RP the axial} $\varepsilon''(\omega)$  peaks.  {\RP Figure \ref{[figure7]} shows $\varepsilon''(\omega)$ for three different armchair SWCNTs, [4,4,m], [14,14,m] and [24,24,m]. }Increasing the SWCNT radius allows more lattice translations in the circumferential direction and thus a larger degree of cutting line packing. Dresselhaus and Saito quantified the density of cutting \cite{[15],[16]} and showed that this increase in the cutting line density means that the cutting line closest to the K point will push closer to it.  
Therefore, the larger the SWCNT radius, the larger its cutting line density and the smaller the difference between its first conduction and valence bands.  This causes a systematic shift in the DOS and $\varepsilon''(\omega)$  peaks to lower energy (a well known effect) \cite{[16], [18], [13]}. Additionally, it means that larger tubes will have more of these peaks because of an increased availability of cutting line transitions.  Note that these shifts of $\varepsilon''(\omega)$ peak position as a function of diameter are only valid for the axial direction $\pi$ peaks.  In {\RP Figure \ref{[figure7]}(a)}, one can see that all the SWCNTs have a final $\varepsilon''(\omega)$ peak (the $\pi_0$ peak) at 4.20 eV in the $\pi$ transition region.  This is due to the cutting lines reaching some common border for all armchair SWCNTs.  All peaks beyond this 4.20 eV $\pi_0$ peak, in the 5-30 eV $\sigma$ transition region in {\RP Figure \ref{[figure7]} (b)} are invariant with change in radius.  

Before investigating the differences in these two regimes, first consider the known variation in $\varepsilon''(\omega)$ peak locations between the metal and semiconductor classifications in the $\pi$ regime.  
When a K point is crossed, the next nearest cutting line is a full spacing away.  Conversely, the semiconducting tubes have their first two cutting lines at 1/3 and 2/3 of this distance. The implications of this are: Metals cross a K-point where the CB and VB meet, allowing for the possibility of a finite-height metal peak all the way down to 0.00 eV.  Symmetry effects are the only thing that prevent it from appearing in the $\varepsilon''(\omega)$ optical dispersion properties.  Additionally, metals have their next nearest cutting line at an energy three times larger than that of a semiconductor of equivalent radius. 
Therefore, while the $\pi$ peaks in $\varepsilon''(\omega)$ from 0-4 eV will shift systematically as a function of radius for tubes of similar electronic structure (i.e., metals or semiconductors), there is a big shift in the peak locations between the classifications \cite{[15], [18]}.

	As seen in {\RP Figure \ref{[figure7]}(b)}, the $\sigma$ transitions seen in the 5-30 eV optical dispersion properties seem to be invariant with respect to changes of CNT radius.  But are the data in this regime the same for all SWCNTs or is there another parameter that can make a difference? If we compare a zigzag semimetal with a zigzag semiconductor, we have essentially fixed the geometry and tested whether there is a correlation with the electronic structure properties of the $\pi$ regime. Likewise, if we compare zigzag semimetals with armchair metals, we have essentially fixed the electronic structure while varying the geometry \footnote{Technically, semimetals and metals have differences in their properties at the Fermi level.  However, we are more interested in whether the relationships of the cutting lines with the K-points also impact the 5-30 eV properties in any significant way, making the crossing/grazing differentiation irrelevant.}.  Comparing all three would allow us to check if there is a simultaneous correlation with both geometry and 0-5 eV electronic structure. {\RP Figure \ref{[figure7]}} (a) and (b) combined with {\RP Figure \ref{[figure9]} (a) and (b)} show just such a comparison of the three SWCNT types for both the 0-5 and 5-30 eV ranges. The $\pi$  regime behaves as expected with its dependencies on radius and zone-folding classification.  The $\varepsilon''(\omega)$ peaks of the metallic and semiconducting zigzag SWCNTs are visually identical in the $\sigma$ regime, eliminating any correlation to the zone-folding classification.  However, there is a clear difference in $\varepsilon''(\omega)$ peak magnitude and position between the armchair and zigzag geometries, suggesting geometry as the influential parameter in the $\sigma$ regime.  

\section{Features and analysis of spectral properties}
\label{sec:features}


	Because we have observed trends in spectral properties for the largest diameter SWCNTs (i.e., in the zone folding regime), it is instructive to ask how resilient they are. Do they work down to the smallest radius SWCNTs?  Is there a particular cutoff for the range/diameter?  Are the departures from the trends subtle or are they large, leading to a seemingly random set of properties among the chiralities?  Figure \ref{[figure10]} (a) to (d) shows $\varepsilon''(\omega)$ peak position vs. radius for the armchair SWCNTs from [3,3,m] to [24,24,m] in both axial and radial directions in two energy ranges separated at the critical radius of 0.7 nm. In the 0-5 eV range there is a smooth trend down to the tiniest of SWCNTs.  However, in the $\sigma$ regime, we see a large distortion of peaks for radii of less than 0.7 nm.  New peaks appear to form and others appear to shift randomly \footnote{Note that a change in quantity of peaks doesn’t necessarily change the overall magnitude.  It simply is a useful parameter to monitor when studying changes among the various chiralities.}.

\begin{figure*}
\begin{center}\includegraphics[width=14cm]{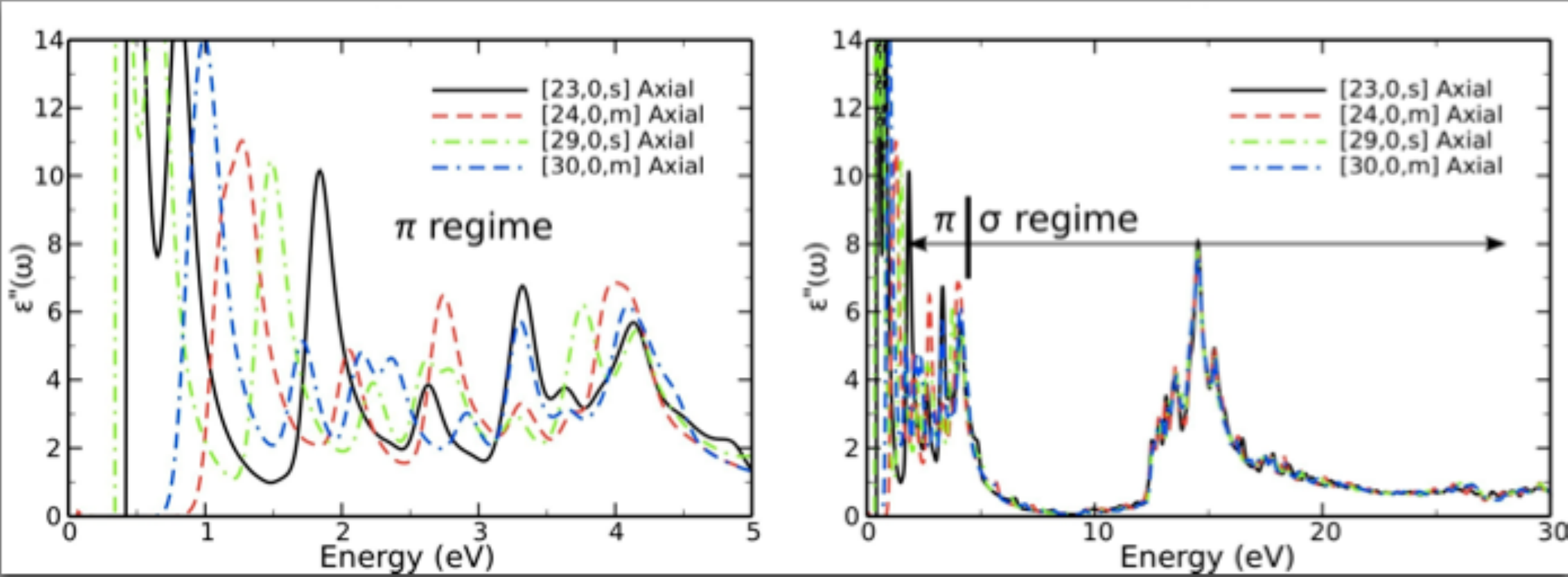}\end{center}
~~~~~~~~~~~~~~~~~~~~~~~~~~~~~~~~~~~~~~~~~~~~~~~~~~~~~~~~~~~~~~(a) ~~~~~~~~~~~~~~~~~~~~~~~~~~~~~~~~~~~~~~~~~~~~~~~~~~~~~~~~~~~~~~~~~~~~~~~~~~~(b)
\caption{Comparison of $\varepsilon"(\omega)$ optical dispersion spectrum in the (a) $\pi$ ( 0-5 eV) and (b) $\sigma$ (5-30 eV) regions along the axial direction for metallic and semiconducting zigzag SWCNTs. Frequency in eV. }
\label{[figure9]}
\end{figure*}


	In the case of SWCNTs we detect additional $\pi-\sigma^*$ and $\sigma-\pi^*$-like transitions due to curvature causing these otherwise independent directions to overlap. As a result, between the limit of 3.8 eV (all $\pi-\pi^*$ transitions below) and 10.0 eV (significant $\sigma - \sigma^*$ transitions above), it is not clear which particular transitions contribute most to the optical dispersion properties.  The goal here, however, is simply to cathegorize the two distinctly different regimes found in Figs. \ref{[figure9]}, and \ref{[figure10]}. Ultimately our selection of the 5 eV as the $\pi/ \sigma$ cutoff was based on three major factors: 1) Large diameter SWCNTs typically have no major $\varepsilon''(\omega)$ contributions between 5-10 eV for the axial and/or radial directions.  Thus we could arbitrarily select any location along this interval without changing around the distributions of peaks contained in each of the groups.  2) Since the edge between $\pi-\pi^*$ and non $\pi-\pi^*$ related $\varepsilon''(\omega)$ contributions is the only hard edge known with certainty, it makes sense to select the lowest energy in the 5-10 eV interval.  Therefore, our $\pi$ classification for SWCNTs is purely $\pi-\pi^*$ based while anything above 5 eV is likely to have contributions from all 4 possibilities ($\pi-\pi^*$, $\sigma-\pi^*$, $\pi-\sigma^*$, and $\sigma-\sigma^*$). 3) In Figure \ref{[figure10]}, the smallest SWCNTs have the largest curvature and therefore the highest degree of $\pi-\sigma^*$ and $\sigma-\pi^*$ overlap.  This is likely the reason for the increase in $\varepsilon''(\omega)$ peaks appearing in the 5-10 eV limit. 

\subsection{Metal Peak Analysis }

\begin{figure*}[!t]
\begin{center}
\includegraphics[width=6.0cm]{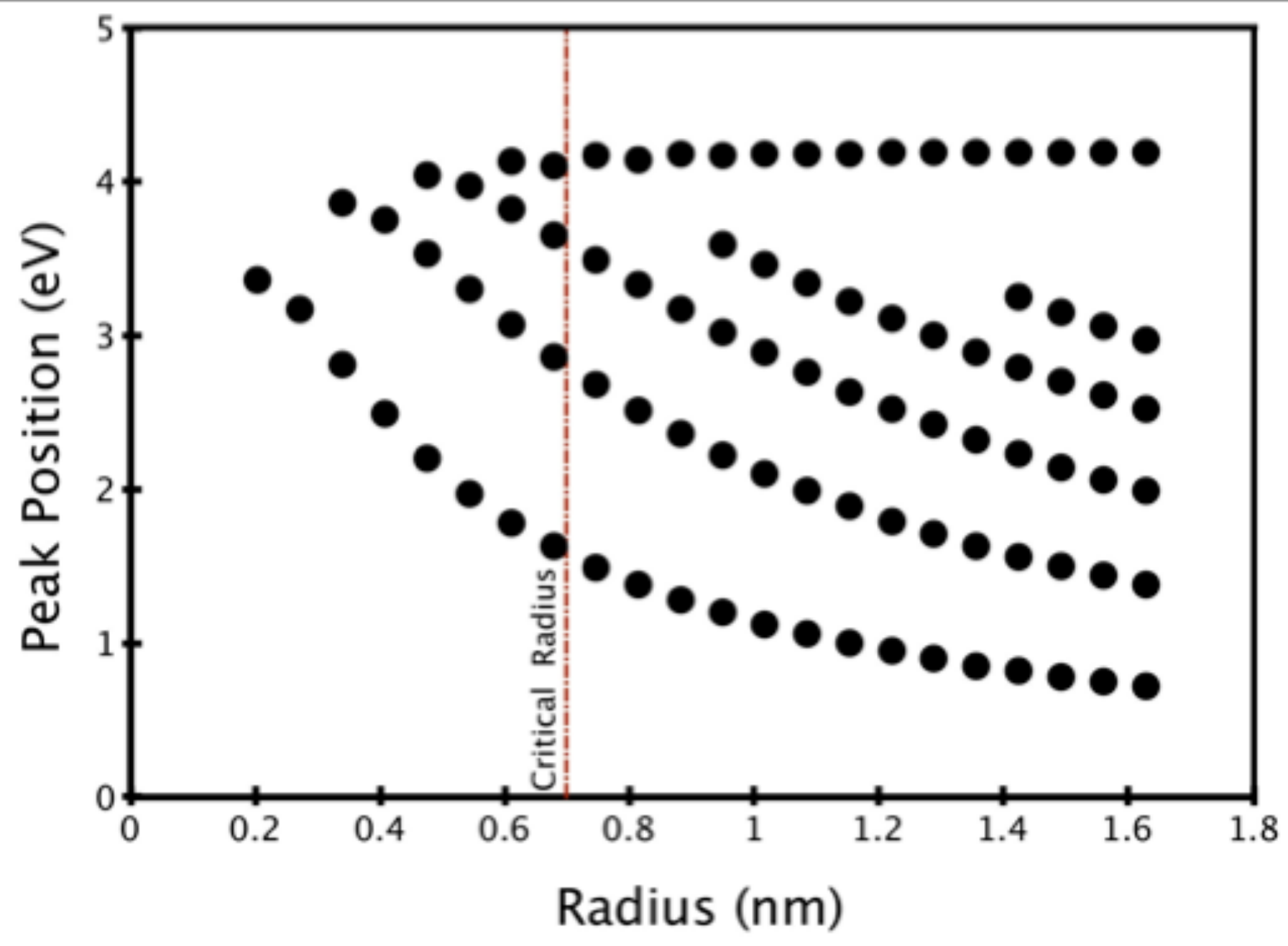}
\includegraphics[width=6.0cm]{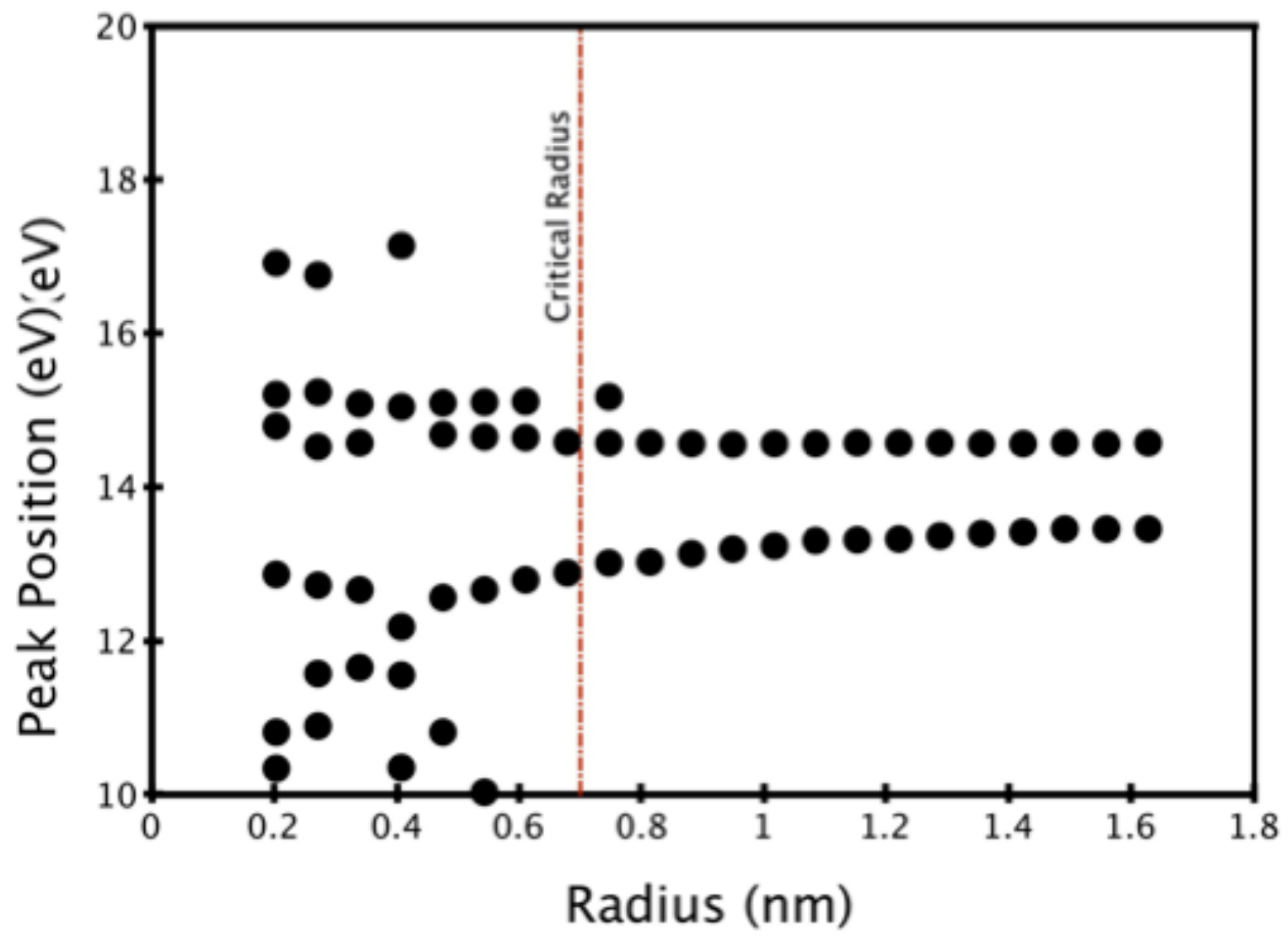} 
\end{center}
~~~~~~~~~~~~~~~~~~~~~~~~~~~~~~~~~~~~~~~~~~~~~~~~~~~~~~~~~~~~~~~~~~(a) ~~~~~~~~~~~~~~~~~~~~~~~~~~~~~~~~~~~~~~~~~~~~~~~~~~~~~~~~~~~~~~~~~~~~(b)
\begin{center}
\includegraphics[width=6.0cm]{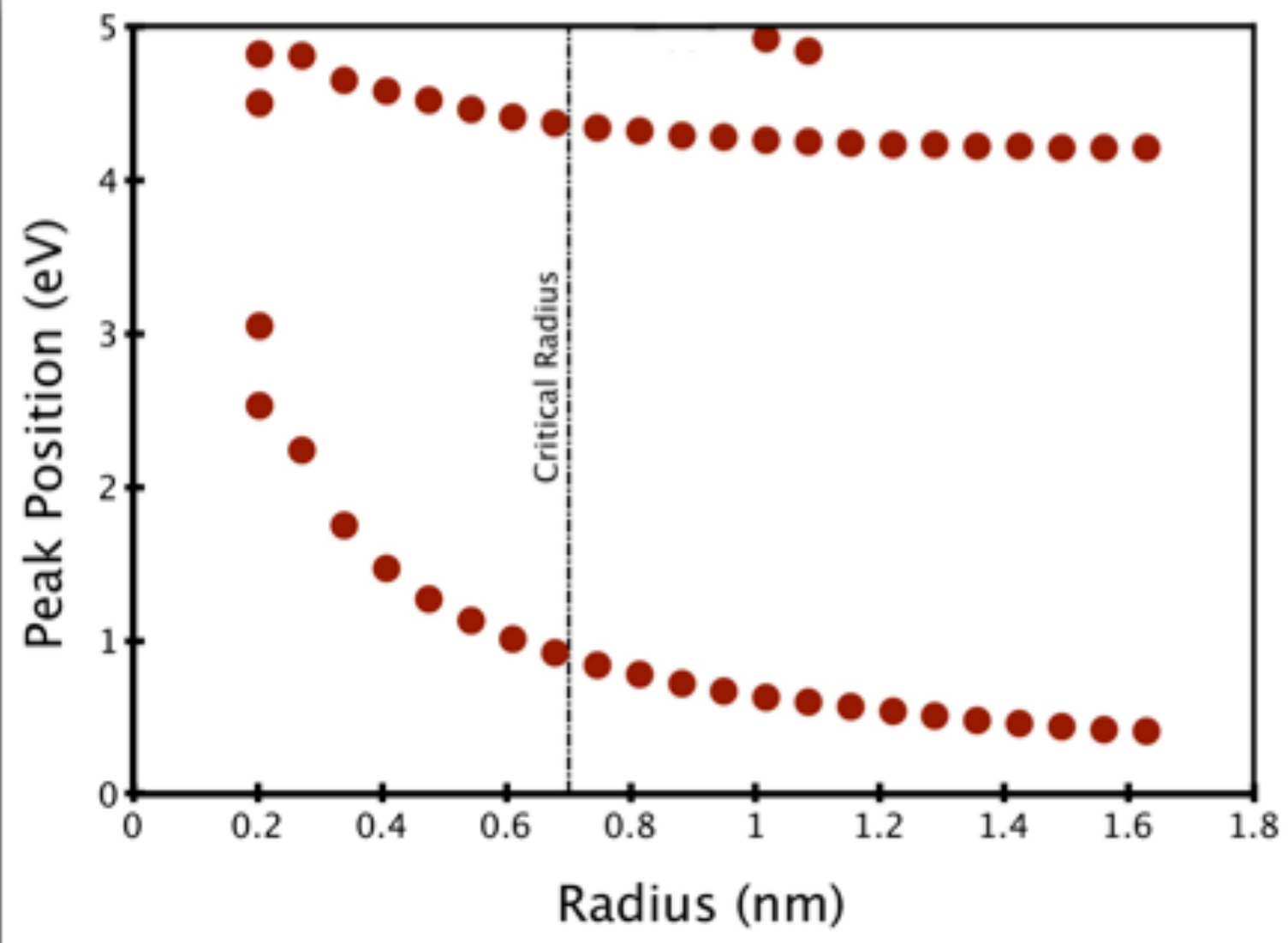}
\includegraphics[width=6.0cm]{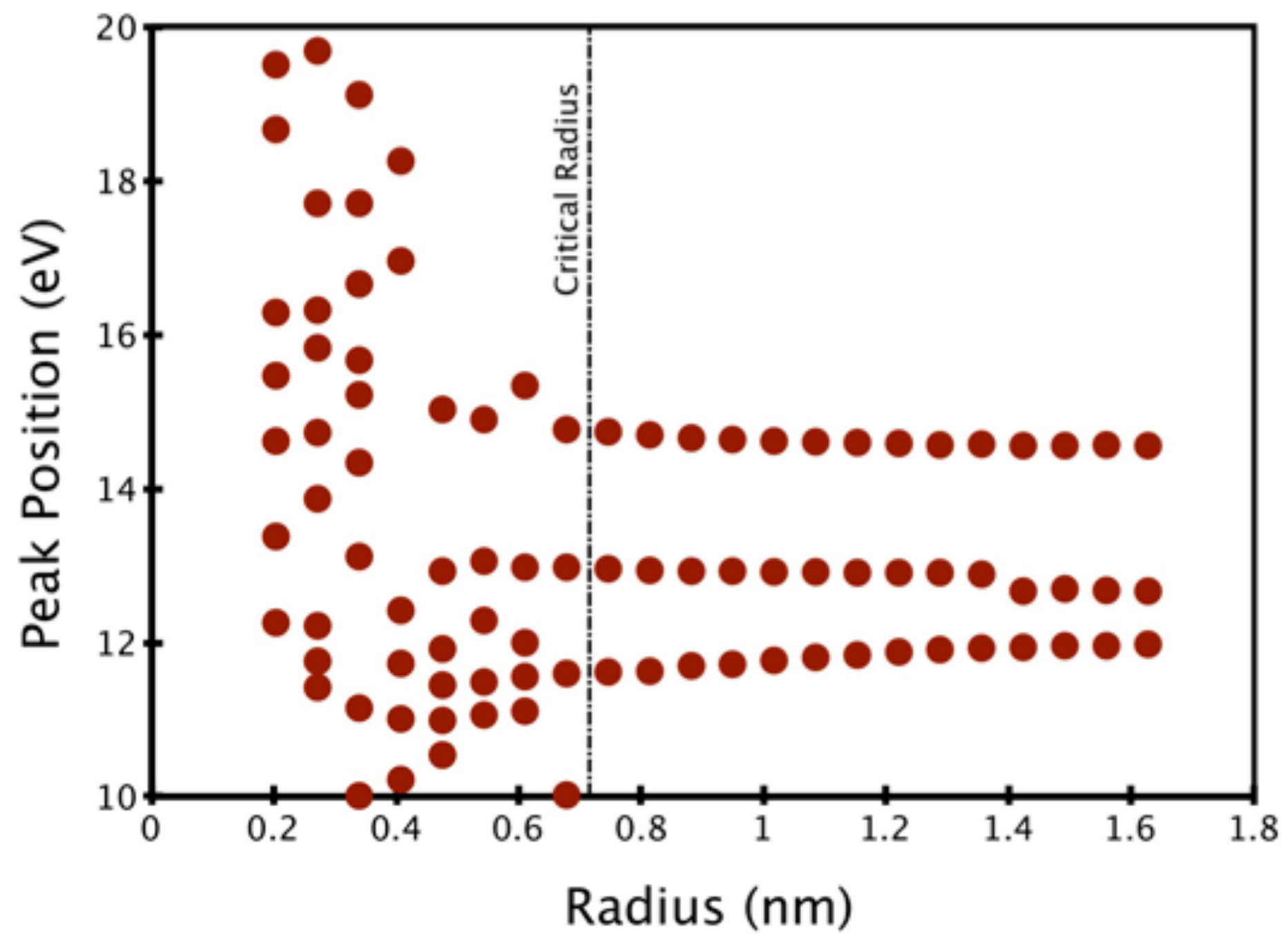}
\end{center}
~~~~~~~~~~~~~~~~~~~~~~~~~~~~~~~~~~~~~~~~~~~~~~~~~~~~~~~~~~~~~~~~~~(c) ~~~~~~~~~~~~~~~~~~~~~~~~~~~~~~~~~~~~~~~~~~~~~~~~~~~~~~~~~~~~~~~~~~~~(d)
\caption{Plots of peaks in optical dispersion spectrum $\varepsilon"(\omega)$ vs. the radius of SWCNTs. (a) axial direction from 0-5 eV in the $\pi$ region. Smooth trends observed for all radii; (b) radial direction from 10-20 eV in the $\sigma$ region. No trends for small radius tubes; (c) axial direction from 0-5 eV in the $\pi$ region. Smooth trends observed for all radii; and (d) radial direction from 10-20 eV. No trends for small radius tubes. Frequency in eV. }
\label{[figure10]}
\end{figure*}

An ideal metal is characterized by a simple Drude model corresponding to 	
\begin{equation}
\varepsilon''(\omega) = \frac{4\pi ~N e^2}{m} \frac{\tau}{\omega (1 + \tau^2\omega^2)} = \omega_p^2 \frac{\tau}{\omega (1 + \tau^2\omega^2)}.
\label{Drude}
\end{equation}
where $N$ is the electron density, $m$ the electron mass, and $\omega_p$ the characteristic plasma frequency. This expresses the divergence of the dielectric response at vanishing frequencies. This feature, while not critical to our studies of SWCNTs, is a topic of interest in the formulation of electrodynamic forces due to the importance of conductivity affects \cite{[58], [59], [60], [61]}.  In computation, we can only approach the static limit asymptotically. 

In computation we can distinguish three types of behavior in this metallic peak energy range. We refer to these behaviors as null, $\rm  M_0$, and $\rm M_1$. Null represents a SWCNT with at least a 0.1 eV optical band gap. For a $\rm  M_0$ peak, $\varepsilon''(\omega)$ rises asymptotically as the energy approaches 0 eV, while a $\rm  M_1$ peak will rise in a similar manner but then falls back to a value of $\varepsilon''(\omega)$ = 0 at an energy in the range from 0.02 to 0.05 eV. It should be noted that very small band gaps (less than 0.05 eV) do exist in materials and in this work the $\rm  M_1$ type metal peaks are not likely due to a computational artifact since we calculate the optical transitions and dipole matrix elements down to 0.01 eV transition energy. 

Consider the optical dispersion properties of three SWCNTs representing the armchair metal, zigzag metal, and chiral metal classes.
In a zone-folding classification, these would all be considered metals and therefore the 0 eV behavior should be indistinguishable for all three cases.  The Lambin classification would state that the [9,9,m] armchair is a true metal, but the [18,0,m] and [9,3,m] would experience a slight gap of less than 0.1 eV. Therefore we would expect to potentially see two types of behavior at 0 eV.  In actuality, we see all three metal behaviors: null (armchair), $\rm  M_1$ (chiral metal), and $\rm  M_0$ (zigzag metal). The most surprising observation is that of the [9,9,m] armchair.  Despite having no band gap and a tiny but continuous DOS around the Fermi level, it exhibits no optical transitions from 0 to 1.5 eV.  This effect is somewhat paradoxical because the armchair SWCNTs represent the only class with no band gap (not even a very small band gap like the zigzag and chiral semi-metals), and yet they exhibit optical band gaps as large as 1 eV. Also unexpected is the $\rm  M_0$ peak of the [18,0,m] because zigzag metals are supposed to have a very small, but finite band gap and thus we would expect an $\rm  M_1$ type peak. Perhaps structural relaxation would resolve this inconsistency and uncover an $\rm  M_1$ type peak instead \cite{[39]}. If, however, this proves to be a real and accurate result, then dividing the metals into these three metal categories is justified because it will result in three types of vdW-Ld  interactions.

\begin{figure*}[t!]
\begin{center}
\includegraphics[width=7.3cm]{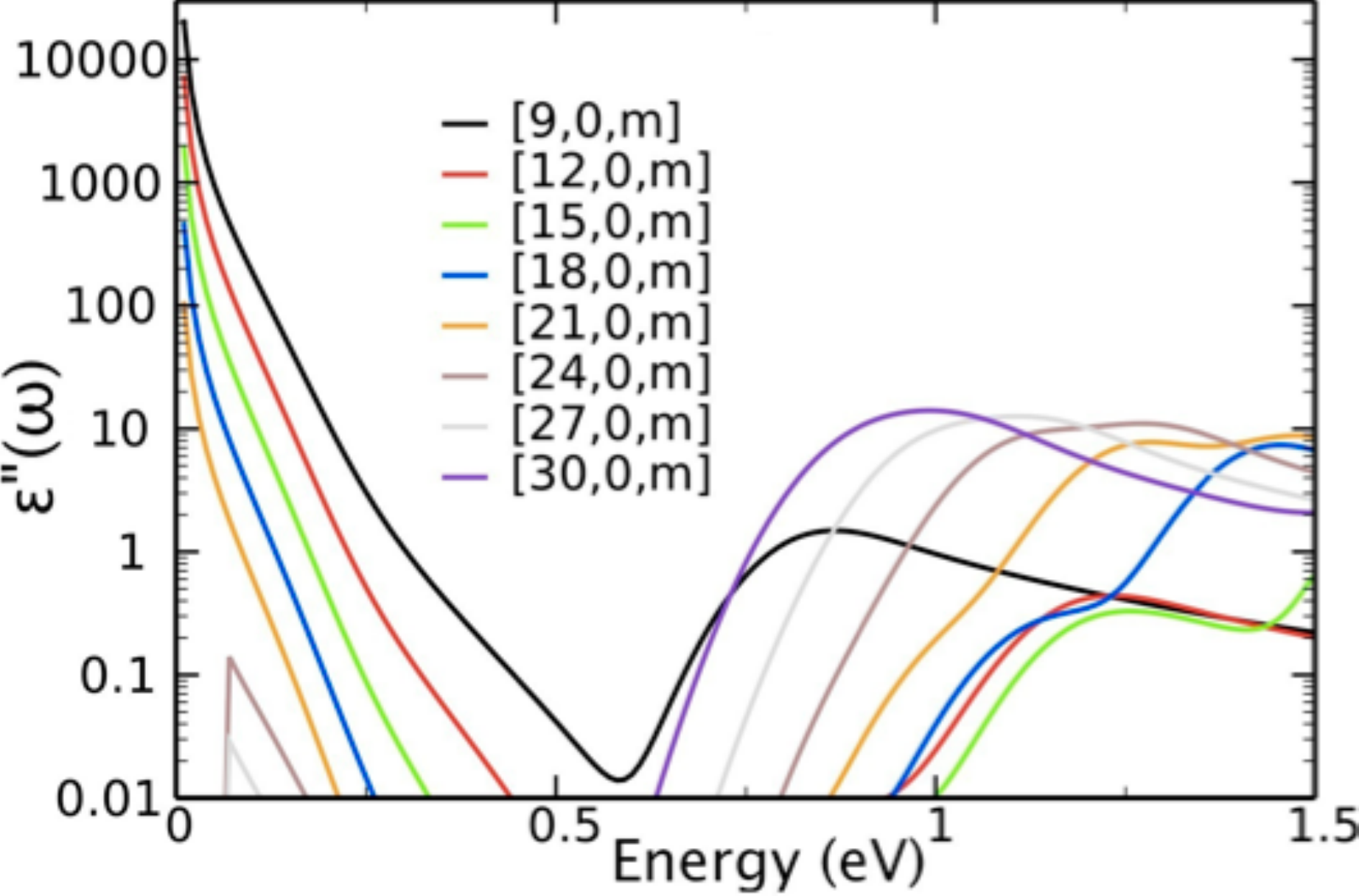}
\includegraphics[width=7.0cm]{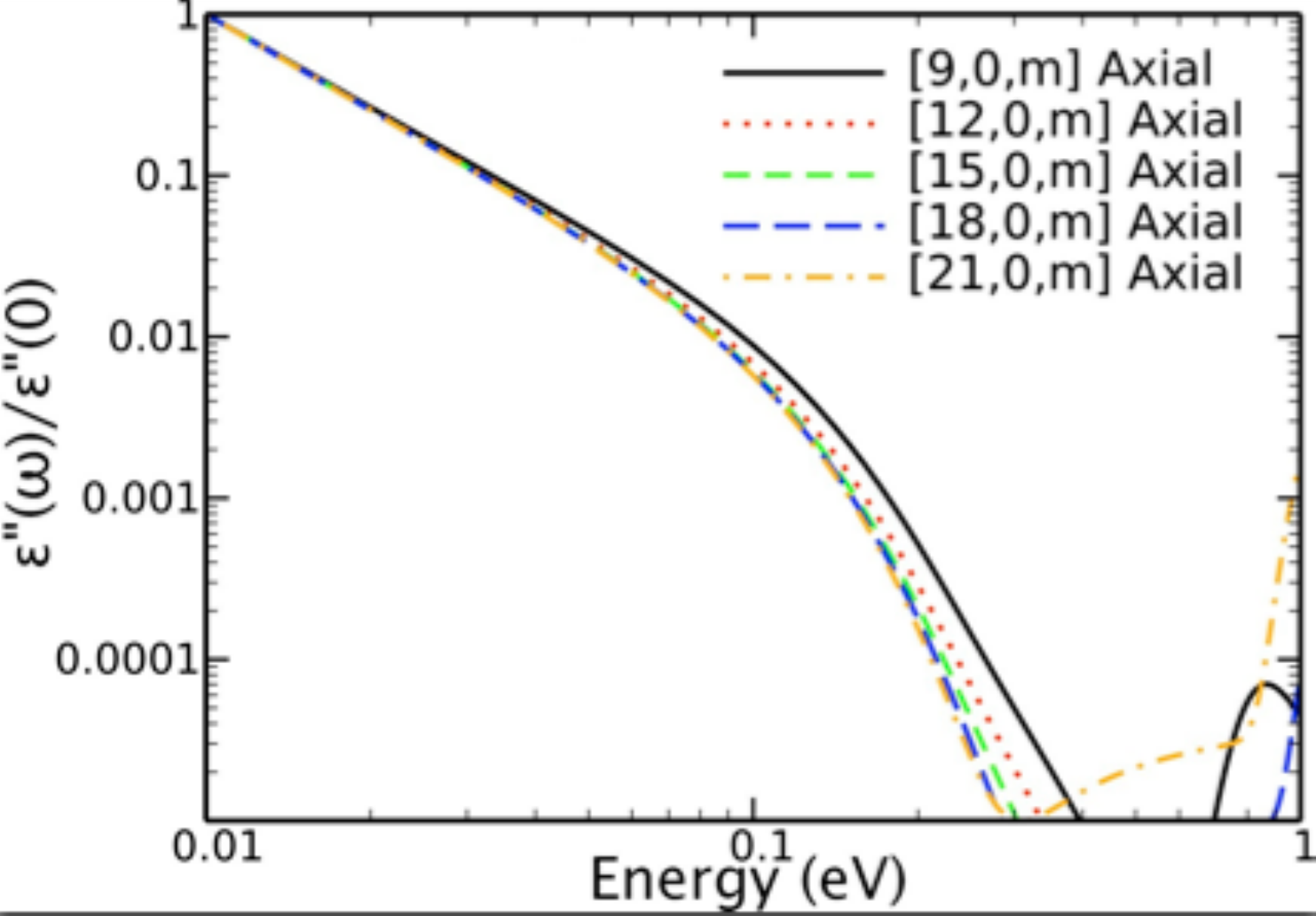} 
\end{center}
~~~~~~~~~~~~~~~~~~~~~~~~~~~~~~~~~~~~~~~~~~~~~~~~~~~~~~~~~~~~~~~~(a) ~~~~~~~~~~~~~~~~~~~~~~~~~~~~~~~~~~~~~~~~~~~~~~~~~~~~~~~~~~~~~~~~~~~~~~~~~~~~~~~(b)
\caption{ (color on line)  (a) $\varepsilon"(\omega)$ optical dispersion spectrum around 0 eV for eight zigzag 
SWCNTs  spanning a large range of diameters. (b) A log-log plot of $\varepsilon"(\omega)$ versus energy for five zigzag metals.  The data below 0.1 eV begins to express a flat slope of -2 , indicating a divergent behavior down to 0 eV. Frequency in eV. }
\label{[figure12]} 
\end{figure*}

	Because physical properties depend strongly on the existence and magnitude of an $\rm  M_0$ or $\rm  M_1$ peak, we wanted to determine if such peaks scale with SWCNT radius.  Figure \ref{[figure12]}(a) compares the $\varepsilon''(\omega)$ dispersion properties in the 0-1.5 eV range across 8 different zigzag metal SWCNTs. There is a clearly inverse relationship between the metal peak magnitude and radius with the [9,0,m] having a $\rm  M_0$ peak of approximately 20,000. This magnitude systematically diminishes until [24,0,m], where the $\rm  M_0$ peak opens up to a very weak $\rm  M_1$ peak. For larger tubes, this peak disappears altogether. 

Do chiral metals experience a similar $\rm  M_0$/$\rm  M_1$ trend as a function of radius? Not according to the metal peaks for the nine chiral metal SWCNTs listed in Table 1. The largest $\rm M_0$ peaks come from the [5,2,m] and the [10,4,m] SWCNTs, which respectively have the smallest and second largest radius within the group.  The other seven chiral metals have $\rm  M_1$ and $\rm  M_0$ peaks that are weaker by 1 to 2 orders of magnitude. More large-diameter chiral metals are need to be studied, but they are more difficult to study as they often require up to an order of magnitude more atoms within the unit cell of the {\sl ab initio} calculation. 

A Drude metal shows a $1/\omega$ divergence of the dielectric response function (Eq. \ref{Drude}), giving a slope of -1 in a log-log plot of $\varepsilon''(\omega)$. Figure \ref{[figure12]}(b) shows this plot for five zigzag metallic single-wall CNTs. Instead of observing a slope of -1 (not shown), we observe a slope of -2. The significance of this difference is not clear .

\subsection{$\pi$ and $\sigma$ region analysis }

All large diameter SWCNTs show a $\pi_0$ peak in  $\varepsilon''(\omega)$ at 4.0-4.3 eV that is invariant with respect to changes in radius for a given SWCNT class and direction (i.e. radial or axial). For armchair SWCNTs [see {\RP Figure \ref{[figure7]}(a) and (b)}] the $\pi_0$ peaks occur at 4.19 eV and 4.21 eV for the axial and radial directions, respectively.  However, for armchair tubes with diameters below the zone-folding limit, these peaks shift noticeably with the axial (radial) direction shifts slightly lower (higher) as a function of 1/radius. Why these peaks are pinned at the large diameter limit and shift in opposite directions at the small diameter limit is unclear.  It is conceivable that all SWCNTs experience some common “edge” in the BZ that begins to dissolve as the cutting line density decreases and the bond distortion increases.  

	The next important peak is associated with the optical band gap, labeled $\pi_1$. All other peaks between $\pi_1$ and $\pi_0$ are incrementally labeled $\pi_2$, $\pi_3$, etc., if they exist. [see Figure \ref{[figure7]}(b)].  The largest SWCNTs in this study have up to six $\pi$ peaks in the axial direction. {\RP  The radial direction typically has only $\pi_0$ and $\pi_1$ peaks with relatively flat, low regions  in between.  


	Zigzag metals and semiconductors have a little more variation in their $\pi_0$ peaks as shown in Figure \ref{[figure9]}(a) and Figure \ref{[figure9]}(b). } The values typically oscillate between 4.00 and 4.30 for the largest zigzag SWCNTs in the axial direction, with the variation increasing versus 1/radius. Just like the armchair SWCNTs, the peaks in the radial and the axial directions shift in the opposite direction as the radius decreases. The $\pi_0$ peaks for chiral metals and semiconductors are a bit varied. We have no data for SWCNT with radius greater than 0.7 nm, or larger than the zone folding limit. It is unclear if they will have a stable $\pi_0$ peak for a certain diameter range. As a whole, $\pi_1 - \pi_5$ peaks trend as expected, particularly above the zone-folding limit.  Below this limit, some $\varepsilon''(\omega)$ peaks (most notably for the chiral geometries) distort and shift in unpredictable ways, underscoring the need to obtain this information for each SWCNT rather than approximate it using the data of another chirality.  In contrast, the armchair class of SWCNTs has $\pi_1 - \pi_5$ peaks that vary smoothly down to the smallest diameters.  The similar geometry, electronic structure, and symmetry probably contribute to this resiliency.

	Like $\pi_0$, the $\sigma$ peaks are typically invariant as a function of radius within a given SWCNT class for diameters above the zone-folding threshold.  Up to three $\sigma$ peaks are typically identified. The $\sigma_0$ peak tends to be around 14.5 with a sister $\sigma_1$ peak around 13 eV. Armchair SWCNT geometries have another more rounded peak around 27 eV whereas the zigzag geometries do not. The remaining peaks that arise for very small diameters are not readily named with this system because they can be sharp, often contain only a limited area under the curve, and are difficult to compare tube to tube.  With a coarser resolution in the $\varepsilon''(\omega)$ data (e.g., 0.05 eV instead of 0.01 eV), many of these peaks simply disappear. We collectively name these $\sigma^*$ because their significance depends on the context of the question to be answered. 

\begin{figure*}
\begin{center}
\includegraphics[width=14cm]{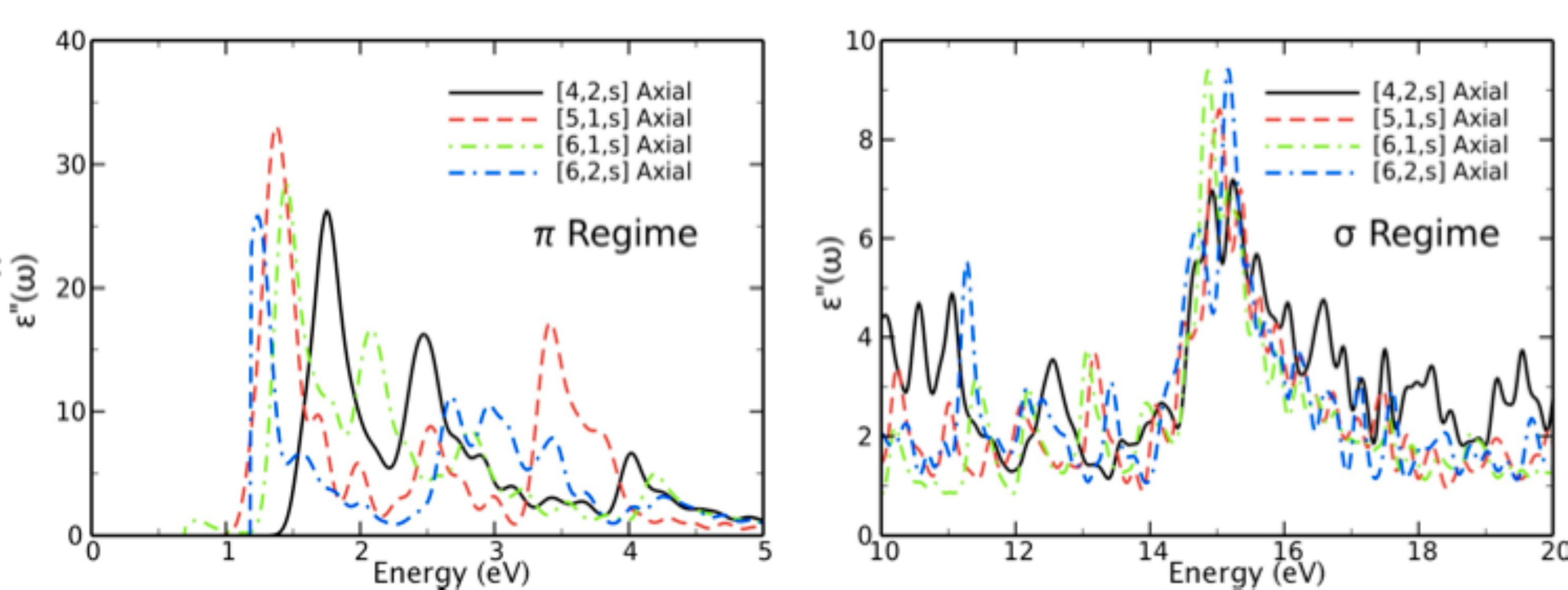}
\end{center}
~~~~~~~~~~~~~~~~~~~~~~~~~~~~~~~~~~~~~~~~~~~~~~~~~~~~~~~~~~~~~~~~(a) ~~~~~~~~~~~~~~~~~~~~~~~~~~~~~~~~~~~~~~~~~~~~~~~~~~~~~~~~~~~~~~~~~~~~~~~~~~~~~~~(b)
\caption{(color on line)  $\varepsilon"(\omega)$ optical dispersion spectrum in the (a) 0-5 and (b) 10-20 eV range for four very small, chiral SWCNTs. Frequency in eV. }
\label{[figure13]} 
\end{figure*}

	For the $\sigma$ peaks, the armchair SWCNTs have values of $\sigma_0$ and $\sigma_1$ of 14.56 and 13.45 (14.56 and 12.69) respectively in the axial (radial) direction.  Thus the location of the $\sigma_0$ peak is the same for both directions and the $\sigma_0$ shifts roughly 0.8 eV. For the zigzag chirality (metal and semiconducting), the $\sigma_0$ peak’s location is also at 14.56 for both directions.  But the $\sigma_1$ peaks correspondingly shift about 0.1 eV to 13.60 and 12.83 for the axial and radial directions, respectively. These shifts may seem small, but changes in position for peaks having large areas under the curve could have a large impact on vdW-Ld interaction.

	Thus far, our primary focus has been on trends of the dispersion properties of a large population of categorized SWCNTs.  At the small diameter limit, there are no stable $\pi_0$ or $\sigma_0$ peaks, and the rest of the peaks can vary in unpredictable ways. Increased curvature introduces more $\rm sp^3$ hybridization, causing the electronic structure to depart strongly from that of graphene \cite{[18]}.  Also, the large decrease in cutting line density means that the allowed states within a given BZ are quite different.  Large optical dispersion variability is found when the different electronic structures are combined with the different allowable samplings. This is illustrated in Figure \ref{[figure13]}, which shows the $\varepsilon''(\omega)$ spectra of four very small diameter chiral tubes in the $\pi$ and $\sigma$ regions. 

\section{General observations on the optical properties of SWCNT}
\label{sec:general_observations}

	Perhaps the most important contribution from the above analysis is the quantification of the variations of $\varepsilon''(\omega)$ in the 0.0-0.1, 0.1-5, and 5-30 eV ranges.  The basic trends that occur in the 0-5 eV range are established in the literature.  The optical dispersion properties in this range are mostly correlated with electronic structure, which depends on the cutting line position (i.e., whether the K-points are crossed) and the packing density (which shifts the peaks toward or away from 0 eV based on the distance of the cutting lines from the K-points).  The only major departure from the previous electronic studies is that of the $\rm M_0$ regime.  Here we discovered that some metals (notably armchair SWCNTs) do not exhibit a standard metal peak despite having the available valence and conduction band states. This results in the {\sl  “metal paradox”} where the armchair “metals” are optically like semiconductors \cite{[40]}.   

	In contrast to the M and $\pi$ regimes, the optical properties in the 5-30 eV range are largely dictated by the effect of underlying geometry on band to band transitions at higher energy. This means all zigzag SWCNTs at the large diameter limit will have identical properties above 5 eV despite having distinct behaviors (i.e., null, $\rm M_1$, or $\rm M_0$ metal peaks and metal versus semiconducting $\pi$ peaks) in the lower energy regimes. 

{\RP Having extended our focus beyond DOS and electronic structure, 
the interesting symmetry considerations} outlined by Dresselhaus, Saito et al. do not necessarily manifest themselves as notable differences in the optical properties \cite{[15]}. Certain optical transitions are symmetry forbidden, unlike the analysis of Katura plots and the VB to CB transition tracking with radius \cite{[62]}.  This is a hazard of stopping at the DOS level analysis and approximating the optical properties without explicitly calculating the dipole matrix elements.  For example, the known differences (metal versus semiconducting position of $\pi_1$) and trends ($\pi_1 - \pi_5$ shifting with radius) lead to easily observable differences in key optical properties, such as the index of refraction.  In the small diameter limit (where the aforementioned trends have broken down) SWCNTs like the [6,2,s], [6,1,s], and [5,1,s] have little correlation in the $\varepsilon''(\omega)$ peaks leading to a distinct set of properties for each of these smaller tubes. Such variations are important for material selection and experimental design. While we cannot make broad predictions without performing calculations, the calculations for vdW-Ld interactions should be straightforward and can be used to find the relevant variations (see below).

	Our new classification is different from the existing ones, both in terms of the particular categories selected and the reasons for their selection.  For example:  Lambin et al., correctly moved beyond the simple zone-folding classification by identifying the armchair metals as true metals while ascribing very small gap semiconductor characteristics to chiral and zigzag metals \cite{[39]}. But 
it is the armchair that behaves optically like 1+eV semiconductors. The zigzag and chiral semimetals with their sub 0.05 eV band gaps exhibit some form of a metal peak. Thus the division between armchair and non-armchair SWCNTs is correct, but the correlation is opposite from expectation. The Dresselhaus/Saito system, despite having interesting geometry and symmetry considerations relating directly to (n,m), appears to be less accurate as a categorization tool for the optical properties in the σ range. As an example, the M-2p metal category in this classification contains both armchairs and chiral metals. But as seen earlier, the metal peaks of these two groups are different and will lead to different vdW-Ld interactions.  This discrepancy is a result of the zone-folding approximation.  Despite these shortcomings in explaining the $\varepsilon''(\omega)$  optical dispersion properties, the descriptors used to create this classification are useful in describing more quantitative differences in cutting line position, angle, and packing density. Whether some $\varepsilon''(\omega)$ optical dispersion properties link with the symmetry differentiation contained with this classification can only be determined when results on more chiral semimetals and semiconductors are available.  Until then, geometry is still a far more important parameter for the 0-30 eV optical properties \cite{[15]}.


\section{van der Waals - London dispersion interactions}
\label{sec:vdw_interactions}

Represented by the chirality vector {\RP (n,m)}, chiral properties of SWCNTs are connected with their vdW-Ld interactions that depend on their optical properties through the dielectric function, $\varepsilon''(\omega)$. The Kramers-Kronig transform of $\varepsilon''(\omega)$ leads to the vdW-Ld spectrum (vdW- Lds) $\varepsilon(\imath\xi)$, i.e., the dielectric response function at the imaginary values of the frequency $\xi$,  that serves as the basic input for the Lifshitz theory of vdW-Ld interactions \cite{[46],[46],[48],Toni,Jalal}. Above, we addressed and discussed in detail the connection between the chirality vector (n,m) and the optical dispersion spectra $\varepsilon''(\omega)$. While non-intuitive features can occur when one tries to categorize the SWCNT properties based on their electronic spectra (e.g. the metal paradox for armchair SWCNTs), most of the relationships between the SWCNT atomic structure and geometry and their optical properties are conceptually straightforward. We now proceed from electronic structure and optical properties of SWCNTs to their vdW- Ld spectra and the ensuing vdW- Ld interactions connecting chiral properties of SWCNTs with long-range interactions.

\subsection{van der Waals - London dispersion spectra}

Connecting the imaginary part of the dielectric function $\varepsilon''(\omega)$ with the corresponding vdW-Ld spectrum, $\varepsilon(\imath\xi)$, the Kramers-Kronig (KK) relation \cite{[61]}
\begin{equation}
\varepsilon(\imath\xi)=1+\frac{2}{\pi}\int_{0}^{\infty}\frac{\omega~\varepsilon''(\omega)}{\omega^{2}+\xi^{2}}d\omega.\label{eq:KK_transform}
\end{equation}
describes the connection between material's response to external electromagnetic fields of a given frequency and the vdW-Ld spectrum, which characterizes the magnitude of spontaneous electromagnetic fluctuations at the frequency $\xi$. In general $\varepsilon(\imath\xi)$ is a real, monotonically decaying function of the imaginary frequency $\xi$  \cite{[61]}. 

Note that the integration in Eq. \ref{eq:KK_transform} is over an infinite frequency range. In practice, it is usually impossible to know the frequency response of a material accurately over the whole frequency range. Fortunately this is unnecessary as long as all the inter-band transition energies determining the optical dispersion spectrum are either known or properly approximated. The $\varepsilon''(\omega)$ at high energies in the optical domain needs to be pronounced in order to contribute to the resulting vdW-Lds in any significant way because of the heavy damping in the denominator of Eq. \ref{eq:KK_transform}. The SWCNT $\varepsilon''(\omega)$ spectra above 30 eV described in this paper were relatively small in magnitude and noisy. Therefore a cutoff energy of 30 eV was selected to best achieve the desired accuracy when evaluating the Kramers-Kronig transforms.

{\RP There exist distinct conterbalancing effects between high energy wide peaks or low energy narrow peaks, and because $\varepsilon''(\omega)$ and $\varepsilon(\imath\xi)$ are connected via an {\sl integral transform}, it is possible to have identical vdW-Ld spectra with very different dielectric functions. Furthermore, excitonic effects that modify $\varepsilon''(\omega)$ with peak structures \cite{Dresselhous2007} will have in general only a small effect on $\varepsilon(\imath\xi)$ \cite{Exciton6}, which remains a smooth monotonically decreasing function of its argument. In fact, as is clear from Eq. \ref{eq:KK_transform}, every peak will contribute to the vdW-Ld transform approximately additively, as the area under the peak, scaling inversely with $\xi^{2}$, {\sl i.e.} inversely with the square of the Matsubara frequency. Since it is the Matsubara frequency summation in the optical regime of the vdW-Ld transform, that determines the Hamaker coefficients and the strength of vdW interactions, see Eq. \ref{pop-stu11} below, the corresponding excitonic effects on these, while of course present, will be by necessity small. This statement is substantiated by a more detailed quantitative analysis based on full measured SWCNT optical spectra containing an explicit excitonic part \cite{Exciton6}.

\subsection{Hamaker coefficient}\label{citesect}

The {\sl Hamaker coefficient} measures } the strength of the vdW-Ld interaction energy between two interacting materials \cite{[61],French-RMP,Bordag}. It is defined with respect to the geometry or configuration of the interaction setup and the shape of the interacting bodies. Its magnitude depends upon the optical contrast between the interacting materials and the bathing medium. It varies as a function of separation,  a consequence of retardation. In the limit of small separations it can be spoken of as a a constant.  In what follows we shall consider only the vdW - Ld interactions at such small separations, either between two parallel SWCNTs or between a single SWCNT parallel to an anisotropic planar substrate \cite{[46],Toni}. 

The cases of larger and very large separations between two SWCNTs and between a SWCNT and a planar substrate have been dealt with in our other publications that also contain full derivations for all separation limits \cite{[46],[46],Toni,Toni2}. Our analysis is strictly applicable only in the limit of infinitely long SWCNTs; for a consistent analysis of finite size effects and ideal metallic static dielectric response one needs to consider an exact general multiple scattering formulation of the vdW - Ld interactions, such as derived by Emig and coworkers \cite{Emig}.

Consider first the vdW - Ld interaction free energies between {\sl two identical infinitely long parallel SWCNTs} at small separations where retardation effects can be neglected \cite{[46]}. We define the dispersion anisotropy parameter for a cylindrical SWCNT, $\gamma^{c}(\omega)$, as
\begin{equation}
\gamma^{c}(\imath\xi)=\frac{{{\varepsilon^{c}}_{\parallel}}(\imath\xi)-{{\varepsilon^{c}}_{\perp}}(\imath\xi)}{{{\varepsilon^{c}}_{\perp}}(\imath\xi)},\end{equation}
where ${{\varepsilon^{c}}_{\parallel,\perp}}(\imath\xi)$ are the parallel (with respect to the cylinder axis) and the perpendicular components of the vdW
- Ld spectrum of the cylinder. The van der Waals interaction free energy \textbf{per unit length} between two such parallel SWCNT of radius $a$, at a separation $\ell \ll a$, has the form \cite{[46]} 
\begin{equation}
g(\ell,\theta;a)=-\frac{\sqrt{a}}{24~\ell^{3/2}}~\left({\cal A}^{(0)}+{\cal A}^{(2)}\right).\label{pop-equ26}
\end{equation} 
where the sum of coefficients  ${\cal A}^{(0)}={\textstyle \frac{3}{2}}k_{B}T~{\cal H}^{(0)}$ and ${\cal A}^{(2)}={\textstyle \frac{3}{2}}k_{B}T~{\cal H}^{(2)}$, is  
\begin{eqnarray} 
{\cal A}^{(0)}+{\cal A}^{(2)} = {\textstyle\frac{3}{2}}~\frac{k_{B}T}{2\pi}{\sum_{n=0}^{\infty}}'\int_{0}^{2\pi}d\psi~ \Delta^{2}\left(\varepsilon^{c}_{\perp}, \varepsilon_{m}, \gamma\right).
\label{pop-stu11}
\end{eqnarray}
with
\begin{equation}
\Delta\left(\varepsilon^{c}_{\perp}, \varepsilon_{m}, \gamma^{c}\right) = \frac{{{\varepsilon^{c}}_{\perp}(\imath\xi_{n})}\sqrt{1+{\gamma^{c}(\imath\xi_{n})} \cos^{2}{\psi}}-\varepsilon_{m}(\imath\xi_{n})}{{{\varepsilon^{c}}_{\perp}(\imath\xi_{n})}\sqrt{1+{\gamma^{c}(\imath\xi_{n})} \cos^{2}{\psi}}+\varepsilon_{m}(\imath\xi_{n}) }
\label{missmatch}
\end{equation}
Here $\varepsilon_{m}(\imath\xi_{n})$ is the vdW - Ld spectrum of the intervening medium. The vdW-Ld response functions in above equations
are given at a discrete set of Matsubara frequencies  $\xi_{n}=\frac{2\pi k_{B}T~n}{\hbar}$, where $k_{B}$ isthe Boltzmann constant and $\hbar$ is the Planck constant divided by $2\pi$. At room temperature, the interval between neighboring Matsubara frequencies is approximately 0.16 eV. The prime in the summation signifies that the first, $n=0$, term is taken with weight ${\textstyle \frac{1}{2}}$.


In general at a finite mutual angle of the dispersion anisotropy axes $\theta$ the vdW interactions are codified by \textsl{two} values of the Hamaker coefficient, i.e. ${\cal A}^{(0)}$ and ${\cal A}^{(2)}$, whereas in the parallel geometry they are given by a single value ${\cal A}^{(0)}+{\cal A}^{(2)}$\cite{[46],[46],Toni,Toni2}.

The most important aspect of the connection between the Hamaker coefficients and the magnitude of the vdW-Ld interactions is the optical contrast, or \textsl{dispersion mismatch} $\Delta\left(\varepsilon^{c}_{\perp}, \varepsilon_{m}, \gamma^{c}\right)$ in Eq. \ref{missmatch}, essentialy proportional to the difference of the dielectric response function over their sum at every abrupt dielectric interface. The Hamaker coefficients for anisotropic materials are thus proportional to the angular ${\psi}$ integral of the product of all the dispersion mismatches in the system. Clearly, if two materials at each dielectric interface at zero mutual angle have identical dielectric responses at a given Matsubara frequency, the mismatch term goes to 0 and there is no contribution to the Hamaker coefficient at that particular frequency. Thus the magnitude of the Hamaker coefficient is dictated by the degree of dispersion contrast at both interfaces.  On the other hand, the sign of the interaction (i.e. attraction vs. repulsion) is determined by the sign of the dispersion mismatch at the two interfaces if they are not identical. Depending on the dielectric response of all the media involved, the vdW-Ld interaction can have both signs \cite{[61],Capasso}.

One also notes that the contributions of the radial and the axial directions are inherently coupled in the Hamaker coefficients. This means that we cannot simply average the interactions of the pure axial ${\varepsilon^{c}}_{\parallel}$ and radial ${\varepsilon^{c}}_{\perp}$ components, particularly if there is a large degree of optical anisotropy in a given material. Optical contrast is thus the root source for vdW-Ld interactions. Geometry determines how that contrast is weighted, while the configuration, or stacking order of the materials and thereby their vdW-Ld spectra in the calculation, determines the sign. There are additional complexities (e.g. retardation \cite{Toni}, presence of many walls in the CNTs, surfactants covering the CNTs \cite{[48]} etc.), but these topics are beyond the scope of this paper and have been treated at length elsewhere \cite{[61]}.

\section{van der Waals - London dispersion spectra}
\label{sec:vdw_spectra}

Of the three types of $\varepsilon''(\omega)$ variations: position, shape and area of the peak, SWCNT vdW-Ld spectra are primarily affected by shifts in $\pi_{1}-\pi_{5}$ dispersion peaks as well as the details of the type of metal peak. In the first part of this paper we defined and outlined these differences in the electronic structure and dielectric response functions in detail. Here we limit ourselves only to the most noteworthy vdW-Ld spectra.

\subsection{Armchair SWCNT vdW - Ld spectra}

VdW - Ld spectra $\varepsilon(\imath\,\xi)$ of armchair SWCNTs are the most straightforward class for several reasons: 1) they have identical cutting line or chiral angles due to their similar geometrical construction, 2) they are all electronic structure metals, 3) they are all null metals, and 4) pragmatically they are by far the easiest to calculate. Points 1, 2, and 3 are advantageous. We can focus solely on the effects of radius before adding additional features (e.g. metal peaks, metal/semiconductor differences in the $\pi_{1}-\pi_{5}$ peak positions, etc). Point 4 becomes an
issue for chiral SWCNTs, which require a larger number of atoms in the band structure calculation and therefore more computation.

\begin{figure*}[t!]
\begin{center}\includegraphics[width=14cm]{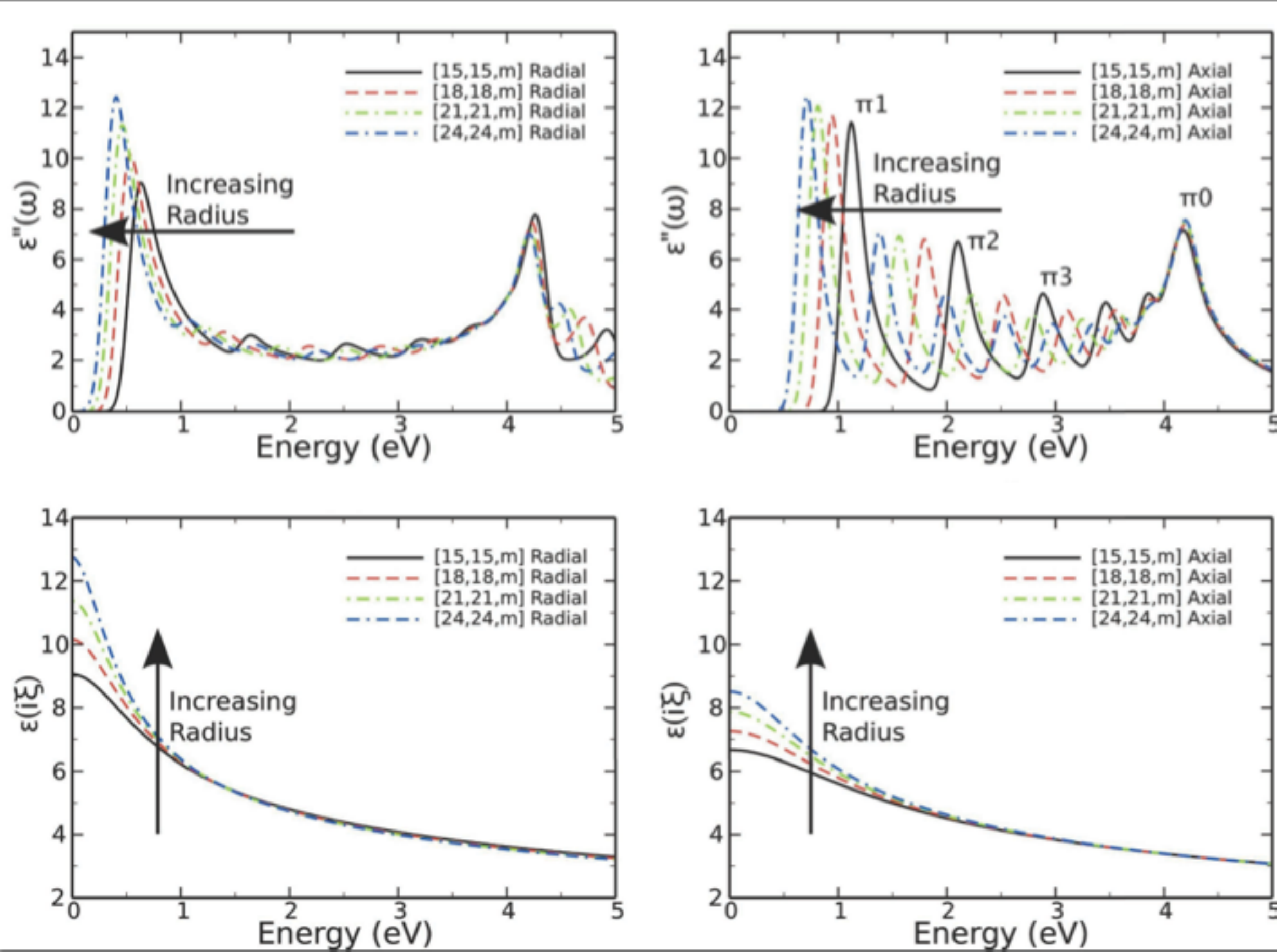} \end{center}
\caption{A comparison of radial and axial $\varepsilon''(\omega)$ (top) and vdW-Ld $\varepsilon(\imath\,\xi)$ (bottom) variations in the radial and axial directions for armchair SWCNTs, from {[}15,15,m{]} to {[}24,24,m{]}. Radii and electronic structure (metal, semimetal, semiconductor) of the SWCNTs are listed in Table 1. Frequency in energy units (eV).}
\label{fig:e2_vdW_armchair_axial_radial} 
\end{figure*}

Large diameter armchair SWCNTs have nearly identical $\sigma$ peaks, no metal peaks, and $\pi_{1}-\pi_{5}$ peaks that shift as a function of inverse diameter. Thus the resulting vdW-Lds are primarily affected by the dispersion properties in the 0.1-5 eV range. Figure \ref{fig:e2_vdW_armchair_axial_radial} compares $\varepsilon''(\omega)$ and $\varepsilon(\imath\,\xi)$ for the armchair axial and radial directions of several large diameter SWCNTs. $\varepsilon''(\omega)$ and the resulting $\varepsilon(\imath\,\xi)$ show smooth and systematic trends. Note the different variation in the high/low energy wings of the dispersion spectrum with respect to radius. While the variation of the radial part of the dispersion spectrum is obviously very different from the variation of the axial response, the final effect on the vdW-Ld spectra is very similar. Both show a general increase in the magnitude of the low energy wing with increasing radius.

There is a clear difference in the behavior of the radial and the axial component of the dispersion spectrum for a given SWCNT, Figure \ref{fig:e2_vdW_armchair_axial_radial}. In our discussion of the optical properties above we outlined the notable difference in the $\pi_{1}$ position between the electronic structure metals and semiconductors. Briefly, the first and subsequent $\pi$ peaks for an electronic structure metal typically occur at a much larger energy than in a semiconductor of equivalent radius. The same is true between the axial and radial directions of the armchair SWCNTs, which behave as electronic structure metals in the axial direction and semiconductors in the radial direction. The result is that semiconducting SWCNTs have their first $\pi$ peak much closer to 0 eV, resulting in a steeper vdW-Ld spectrum around 0 eV.

As shown above, the regularity in the trends of dispersion peaks with respect to radius began to break down at SWCNT radii of approximately 0.7 nm (e.g. the zone-folding limit). However this breakdown in behavior of dispersion peaks does not entail a corresponding change in the vdw-Ld spectra for smaller radii tubes. Small changes in the shape of $\varepsilon''(\omega)$ are not enough to impact the vdW-Lds. Figure \ref{fig:e2_vdW_armchair_axial_radial} compares the resulting axial and radial direction vdW-Ld spectra for the armchair SWCNTs ranging from the {[}24,24,m{]} all the way down to the {[}3,3,m{]}. These results show that the trends are systematic down to {[}5,5,m{]}, which has a radius of 0.34 nm or about half the breakdown diameter size for the dispersion peaks. Therefore, we can expect that other SWCNT classes will exhibit similar resiliency of the systematic trends in the behavior of vdW-Lds for smaller diameter SWCNTs. However, SWCNT types that are less symmetric (i.e. chiral semimetals and semiconductors) are less likely to have such clean trends; the larger variation in their interband transition peaks in the dielectric function is caused by a more diverse set of atomic structures.

\subsection{Metal SWCNT Ld spectra}

Metal peaks can contribute strongly to vdW-Ld spectra, Hamaker coefficients, and consequently to total vdW-Ld interaction energies. According to the classification introduced in the first part of this paper, we group the three types of behavior in the dielectric function that were observed in the 0.00-0.10 eV metal regime into null, M0, and M1 classes. 
Despite this, SWCNTs having a continuous DOS through the Fermi level, the armchair SWCNTs have no metal features while the chiral metals exhibit both M0 and M1 metal type peaks. This is the {\sl metal paradox} whereby symmetry forbids certain interband transitions, giving them a vanishing dipole matrix element despite the available conduction to valence band states.


The divergence of axial vdW-Ld spectra near $\xi=0$ depends strongly on both the size and location of the interband transition peaks. Thus, for a given peak magnitude and area under the curve, an M0 peak would result in a larger vdW-Ld spectral amplitudes since its peak in the dielectric function is shifted to 0 eV. However, an M1 peak can still have a larger effect than a M0 peak if its oscillator strenght is significantly greater.
One thus cannot directly correlate one dielectric function peak with one type of vdW-Ld spectral feature because the integral transform between the dielectric function and the vdW-Ld spectrum. Yet there is a tendency for the metal M0 peaks to contribute more strongly for a given peak height and area under the peak.

\begin{figure}[h!]
\includegraphics[width=8cm]{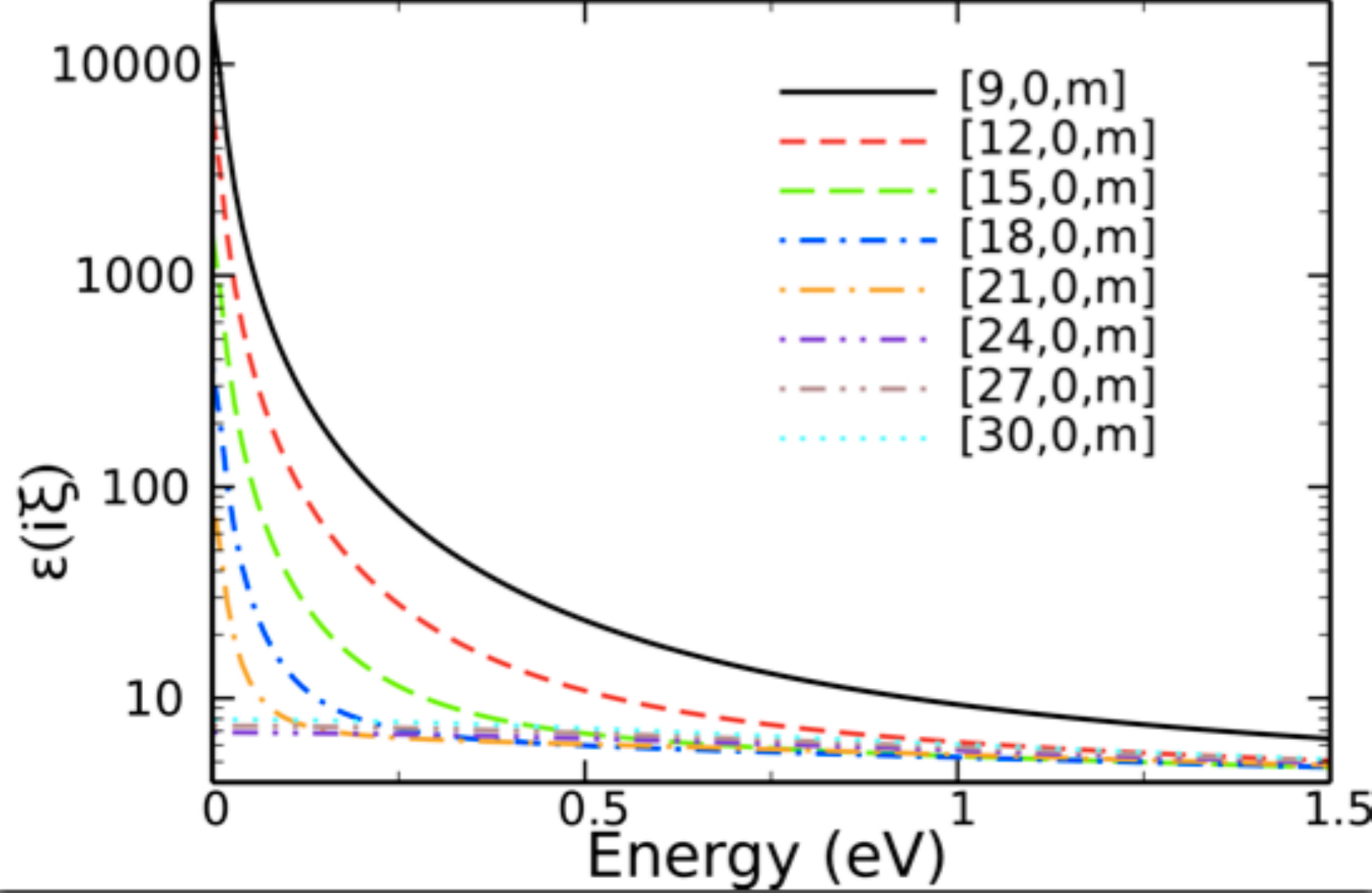} 
\caption{Axial $\varepsilon(i\xi)$ of zig-zag metals for energies close to 0 eV.  The vdW-Ld spectrum $\varepsilon(i\xi)$ is obtained from the appropriate $\varepsilon''(\omega)$ presented in Fig. \ref{fig:e2_vdW_armchair_axial_radial}. Frequency in eV.} \label{fig:e2_LD_peaks_zigzag_zoom} 
\end{figure}

The M0 and M1 axial peaks can sometimes vary systematically as a function of radius within a given SWCNT type. In the first part of this paper, the zigzag type of SWCNTs are found to have M0 peaks for the {[}9,0,m{]} all the way to the {[}21,0,m{]}, varying inversely with radius. For zigzag metals larger than the {[}21,0,m{]}, the M0 peaks give rise to a M1 peak with a magnitude that eventually became negligible by the {[}30,0,m{]} SWCNT,  Figure \ref{fig:e2_LD_peaks_zigzag_zoom}.

\begin{figure*}[!t]
\begin{center}
\includegraphics[width=14cm]{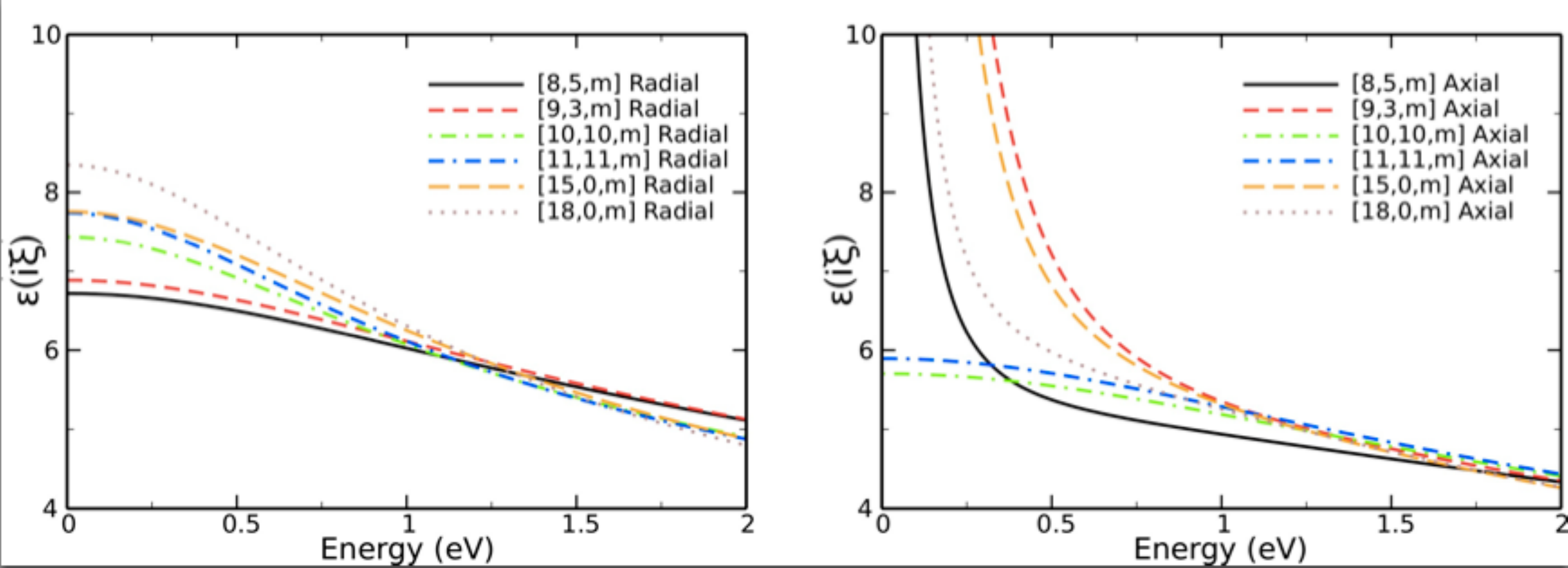} 
\end{center}
\caption{The radial (left) and axial (right) vdW-Ld spectra of SWCNTs representing all three structural types of electronic structure metals: armchair, zig-zag, and chiral. Examples of radial and axial vdW-Ld spectra for similar diameter SWCNTs within each structural class. For details see Table 1. Armchair SWCNT {[}10,10,m{]}, {[}11,11,m{]}, zigzag SWCNT {[}15,0,m{]}, {[}18,0,m{]}, and chiral SWCNT {[}8,5,m{]}, {[}9,3,m{]}.  Frequency in eV.}
\label{fig:metal_compare} 
\end{figure*}

In comparison, chiral metals lack the systematic shift in M0 and M1 peaks as a function of radius; consequently their vdW-Lds magnitudes are harder to classify. Additionally, the differences in their $\sigma$ and $\pi$ peak positions as a function of their chiral angles result in further changes of spectral features across the entire energy range. Thus the armchair, zigzag, and chiral metals can offer very diverse properties. Figure \ref{fig:metal_compare} shows specific examples of radial and axial vdW-Ld spectra for similar diameter SWCNTs within each structural class.

\subsection{Semiconductor SWCNT Ld spectra}

\begin{figure*}[!t]
 \begin{center}\includegraphics[width=14cm]{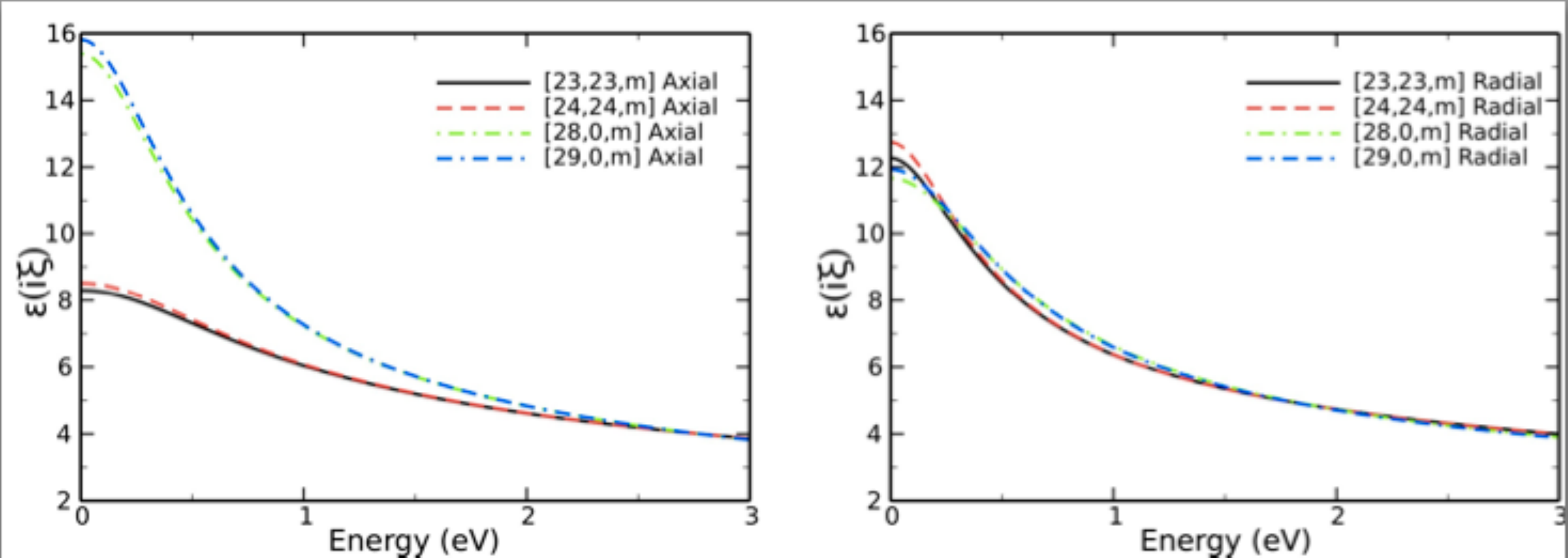} 
 \end{center}
 \caption{The axial (left) and radial (right) vdW-Ld spectra for the armchair ({[}23,23,m{]},  {[}24,24,m{]}) and electronic structure semiconducting zig-zag ({[}28,0,m{]},  {[}29,0,m{]}) SWCNTs. Radii and electronic structures from Table 1. Note that the vdW.Ld spectra of different types of SWCNTs can cross. Frequency in eV.}
\label{fig:LD_armchair_vs_zigzag_semi} 
\end{figure*}

Large-diameter zigzag semiconductor SWCNTs are the last of the three classes of SWCNTs in which we can observe the effects of increasing radius while fixing the chiral angle and thus limiting variability in the $\sigma$ interband transition peaks. Obtaining optical properties such as the dielectric function for very large diameter chiral SWCNTs is still difficult but will likely become easier with greater computational resources. Figure \ref{fig:LD_armchair_vs_zigzag_semi} compares vdW- Ld spectra for large zigzag semiconductor SWCNTs with those of the armchair metal SWCNTs. Because these SWCNTs are semiconducting in the axial direction, their $\pi_{1}$ peak occurs much closer to 0eV and results in a sharper vdW-Lds wing near 0 eV. Because of the lack of a metal peak and a shift of $\pi_{1}$ peak out to higher energies, the armchair metals have a vdW-Lds slope that is comparatively much flatter. However, the behaviors of the radial vdW-Ld spectra for these two geometrical types are similar because their $\pi$ dispersion properties are largely unaffected by the relationship of the cutting lines with the K-point crossings. Variations in the $\sigma$ properties are identifiable but may be ineffective as  a change in shape versus a change in dispersion spectrum peak-width or position.

In the armchair classification, variations in the dielectric function appeared to cancel and result in smoothly varying vdW-Lds down to the smallest SWCNTs. However, small diameter chiral SWCNTs were found to have considerably more spectral variation because of the changing electronic structure and cutting line angles. The result is a collection of vdW-Lds that don't appear to behave in any systematic or definable way. Figure \ref{fig:e2_Lds_small_fries} compares $\varepsilon''(\omega)$ and $\varepsilon(i\xi)$ of armchair and chiral SWCNTs of varying electronic structure properties. Any correlations are inconsistent and unpredictable. The {[}6,2,s{]} and {[}6,4,s{]} have nearly identical axial direction $\varepsilon(i\xi)$, but differ greatly from {[}6,1,s{]}, which has an axial $\varepsilon(i\xi)$ similar to that of {[}5,1,s{]}. While a given pair may have similar axial $\varepsilon(i\xi)$, they might still have completely different radial $\varepsilon(i\xi)$, and vice versa.

\section{Hamaker Coefficients}
\label{sec:hamaker}

\begin{figure}[t!]
\includegraphics[width=7cm]{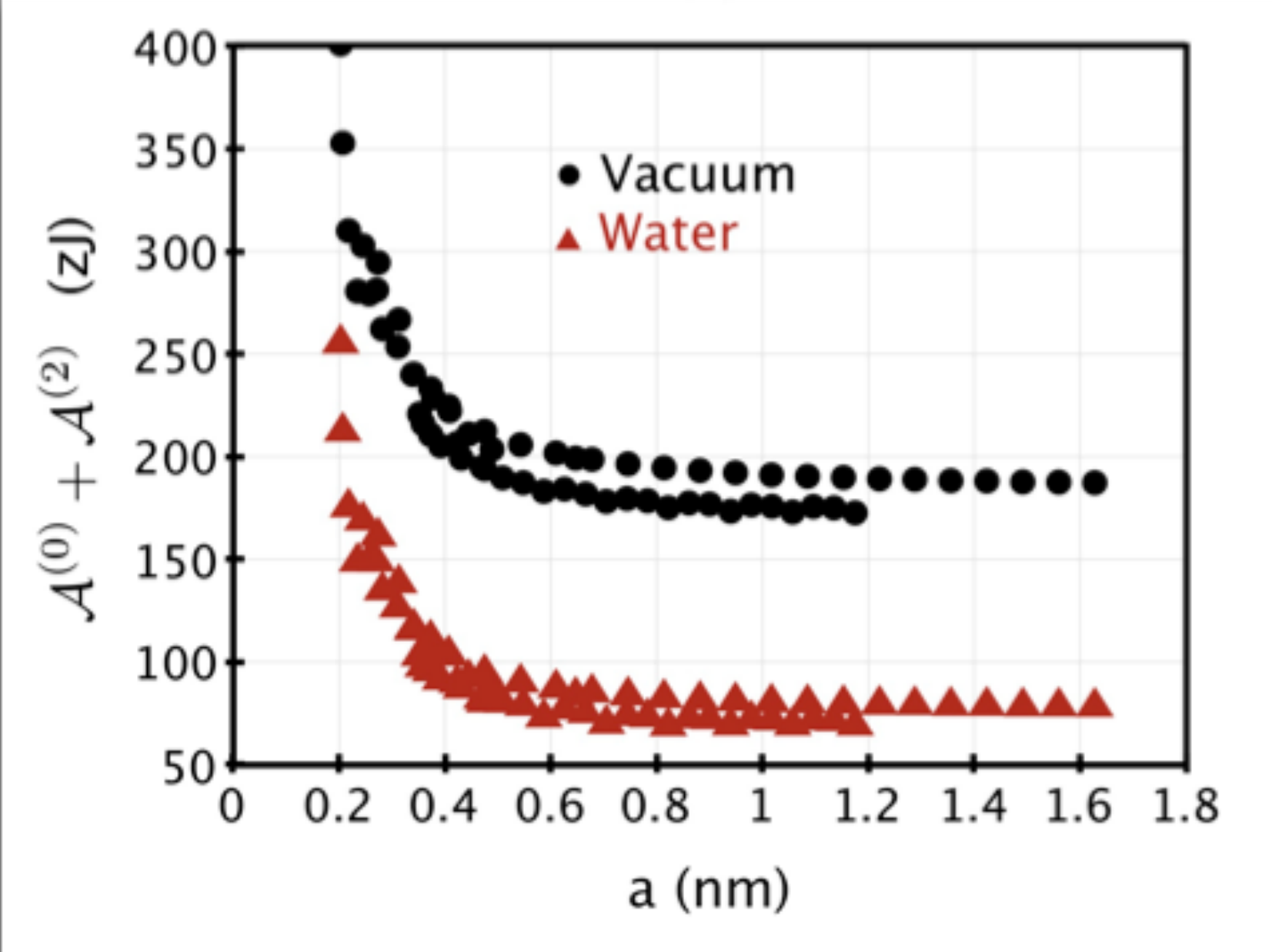}
\includegraphics[width=7cm]{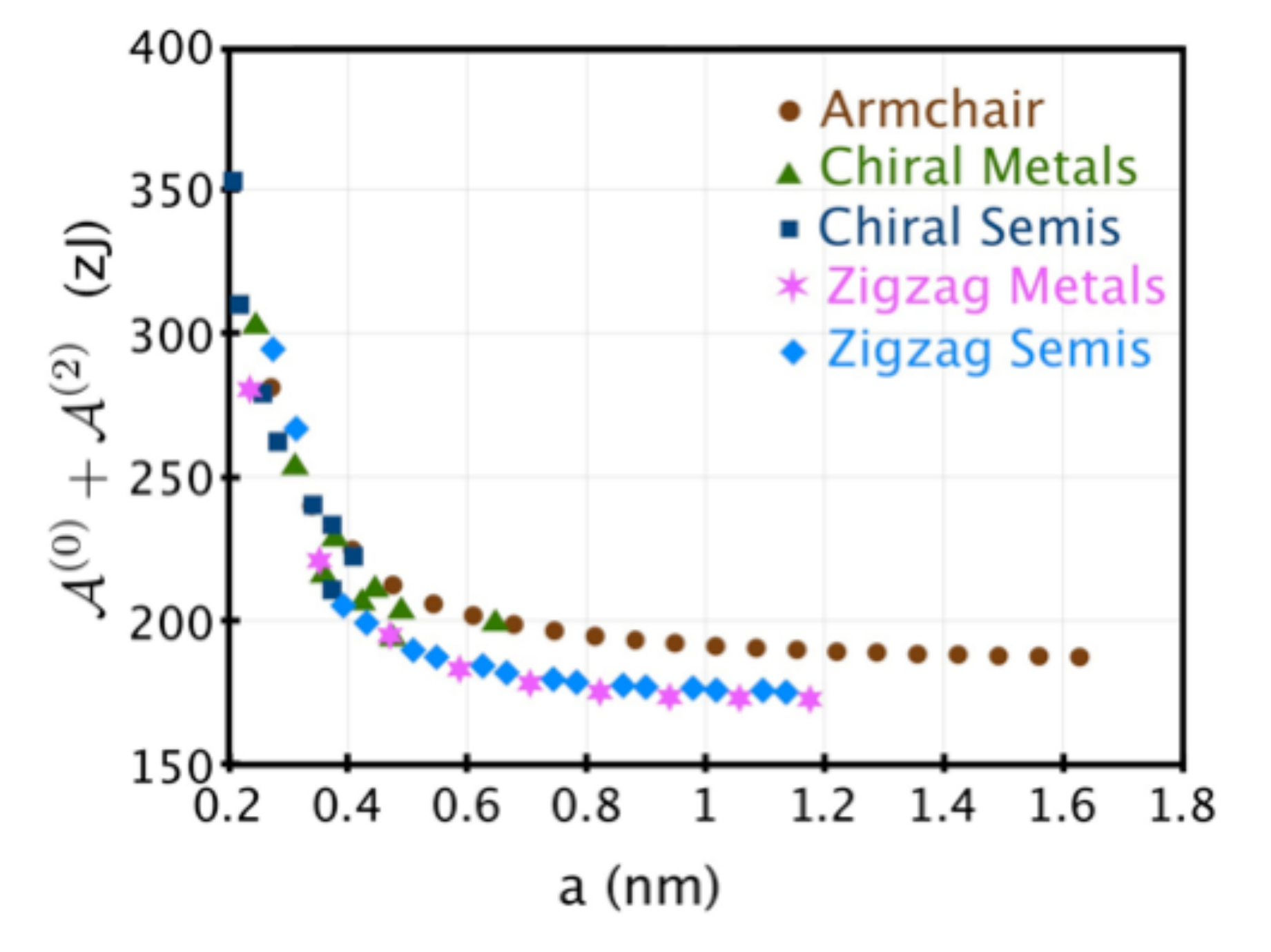} 
\caption{Comparison of Hamaker coefficients, Eqs. \ref{pop-stu11}, vs. SWCNT radius $a(nm)$ for a cylinder-cylinder geometry at small separations, in vacuum and water. Top:  All calculated SWCNTs in vacuum vs. in water. Bottom:  Comparing the 5 vdW-Ld classifications {\sl in vacuo}. The water dispersion spectrum is canonical, taken from Ref. \cite{[61]}.}
\label{fig:A_hollow_rod-rod_near_water_vac_vdw} 
\end{figure}

\begin{figure*}[t!]
\begin{center}
\includegraphics[width=13cm]{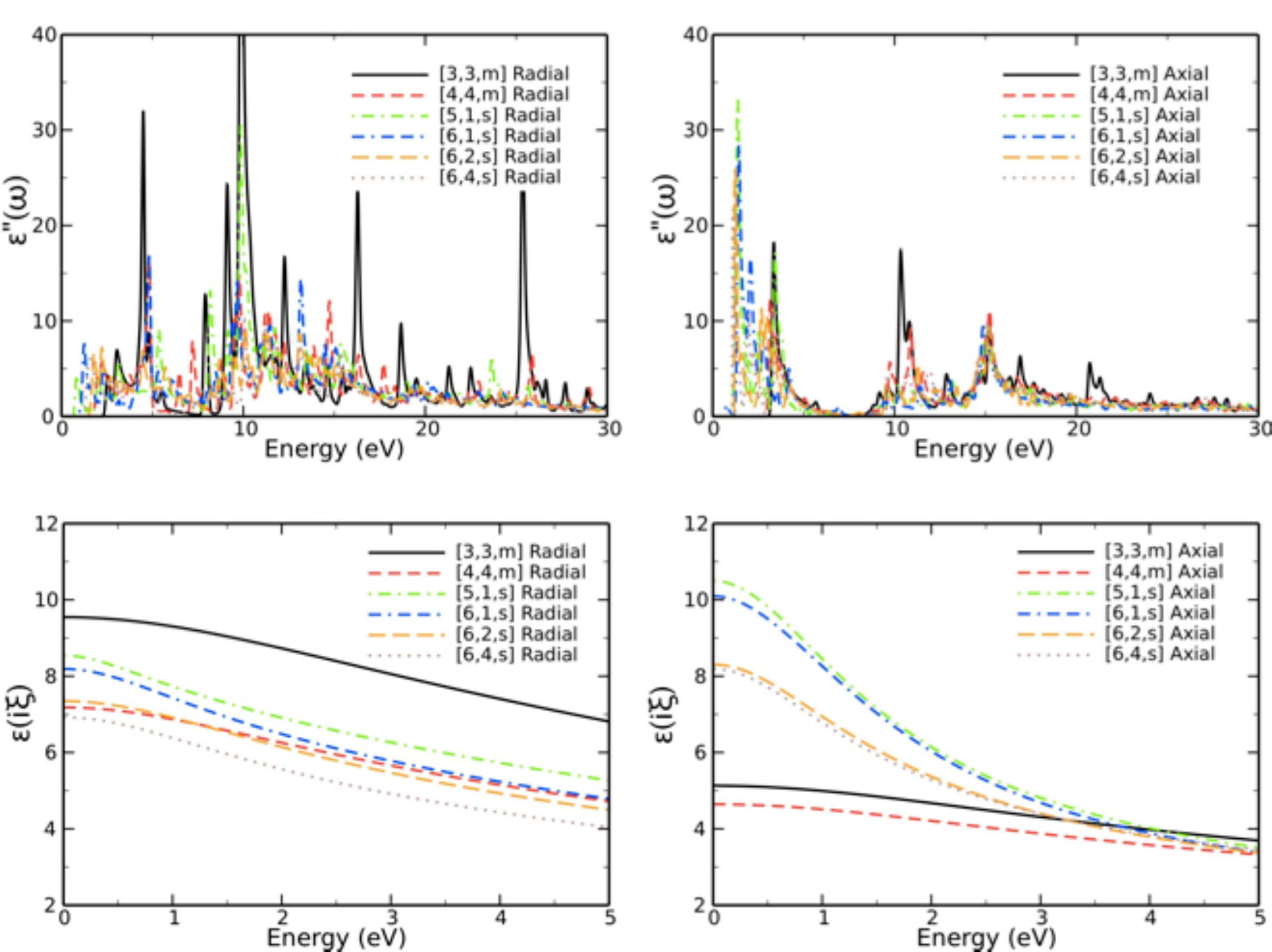}
\end{center}
\caption{{\RP A comparison of radial and axial $\varepsilon''(\omega)$ (top) and $\varepsilon(i\xi)$ (bottom) for very small diameter armchair  ({[}3,3,m{]},  {[}4,2,m{]})  and chiral  ({[}5,1,s{]},  {[}6,1,s{]}, {[}6,2,s{]}, {[}6,4,s{]})  electronic structure semiconducting SWCNTs}. The radii and electronic structure of the SWCNTs are listed in Table 1. Frequency in eV.}
\label{fig:e2_Lds_small_fries} 
\end{figure*}

Do these trends in the electronic structure and corresponding vdW-Ld spectra persist to the level of the Hamaker coefficient and vdW interactions? Figure \ref{fig:A_hollow_rod-rod_near_water_vac_vdw} (a) shows the ${\cal A}^{(0)}+{\cal A}^{(2)}$ Hamaker coefficients derived in Sec. \ref{citesect}, at small separations for cylinder-cylinder interactions for all 64 SWCNTs across vacuum and water, using  Eq. \ref{pop-stu11}. 
Intricate and systematic variations motivate further examination.

In the limit of large CNT radii, the Hamaker coefficients for different classifications of SWCNTs trail different asymptotes, Figure \ref{fig:A_hollow_rod-rod_near_water_vac_vdw}. The numerical values of the Hamaker coefficients are relatively flat and unchanging above the zone-folding limit corresponding to a CNT radius of about 0.7 nm. Below this limit, their magnitudes begin to shift upward as an inverse function of diameter. One observes a widening envelope that encompasses all the variations in the Hamaker coefficients and becomes larger as the tube radii get smaller. For the smallest tubes, the difference in the Hamaker coefficients can be almost an order of magnitude between different classification types.

The large variation in vdW-Ld spectra as a function of {\RP (n,m)} results in a substantial variation in the corresponding Hamaker coefficients, a result that could not have been predicted by the tight binding approximation, electronic structures classification analysis, force-field modeling, or other approaches other than the explicit determination of the electronic structure and optical properties and the resulting full spectral Hamaker coefficients.

Hamaker coefficients and vdW-Ld interactions show a wide variation among different CNT types.  To simplify we limit the discussion here to the particular exemple of two parallel cylindrical SWCNTs at the small separation limit (see \cite{[46],[48]}).  A system design example will be included in the remaining two subsections to demonstrate practical usage (See also the {\sl Gecko Hamaker project}, \url{http://sourceforge.net/projects/geckopro}, a free and open source project that enables the calculation of full spectral Hamaker coefficients for these and many other systems).

\subsection{Hamaker Coefficients and SWCNT Properties}

Figure \ref{fig:A_hollow_rod-rod_near_water_vac_vdw} (b) compare Hamaker coefficients for various classes: armchair, chiral metals, chiral semimetals, zigzag metals and zigzag semimetals. The Hamaker coefficient behaves systematically across all individual classes but exhibits substantial variation between classes. This is not surprising given the variety of metal and $\pi$ dispersion peak behavior among the different classes. The larger diameter tubes in Figure \ref{fig:A_hollow_rod-rod_near_water_vac_vdw} (b) have a clear ranking in Hamaker coefficient strength as a function of spectral class. However, these trends begin to cross over below the 0.7 nm zone folding limit, and there is also some additional noise/variation from the introduction of the chiral classes. Overall, Hamaker coefficient strengths for the SWCNTs drop by a factor of 2 when the radius decreases from the 0.2 nm to 0.5 nm. For tubes larger than 0.5 nm, the decrease in Hamaker coefficient continues through the largest tubes examined here. 

We cannot completely attribute the larger range of Hamaker coefficient magnitudes in Figure \ref{fig:A_hollow_rod-rod_near_water_vac_vdw} to the effects of increasing curvature at the small diameter limit. The large diameter limit represents only 3 of the 5 SWCNT categories because of the computational difficulty in obtaining very large diameter chiral metals and semiconductors. It is possible that the variation at this limit will be seen once these last two types are included at this diameter limit.

The magnitudes of Hamaker coefficients for larger radius SWCNTs in no way prevents the ability to separate these SWCNTs.  To separate by class, one would design methods that target the different asymptotic limit behavior of the Hamaker coefficients. To separate within a class (perhaps by radius) one would have to create a Hamaker coefficient that would divide along the classification, see below.

\subsection{Hamaker Coefficients and the Interaction Medium}

In Figures \ref{fig:A_hollow_rod-rod_near_water_vac_vdw}(a)  we show the corresponding Hamaker coefficients in water and {\sl in vacuo}. For all but the smallest SWCNTs, the Hamaker coefficients differ systematically by a factor of 2 between these two media.

\subsection{Hamaker Coefficients and SWCNT Size}

For macroscopic systems, material properties are typically independent of size. However, for small scales the material properties and size can be coupled, as is the case here for SWCNTs whose dispersion spectra are connected with their radii. The same is true at very large diameters, where vdW-Ld spectra and Hamaker coefficients can be again very similar.

For example, the 1\% drop in the Hamaker coefficient between the {[}18,18,m{]} and {[}24,24,m{]} SWCNTs is small relative to the 33\% increase in the radius $a$. However,  the vdW-Ld interaction energy varies as $\sqrt{a}$, Eq. \ref{pop-equ26}. This means the {[}18,18,m{]} and {[}24,24,m{]} may have a Hamaker coefficient variation of only 1\%, but the total energy varies by 13\%. In short, Hamaker coefficients can act as useful guides to gauge the interaction strength but they cannot be the sole source for calculating the total vdW-Ld interaction energy that contains explicit dependence on the radius of the tube.

\begin{figure}[!t]
\includegraphics[width=7.0cm]{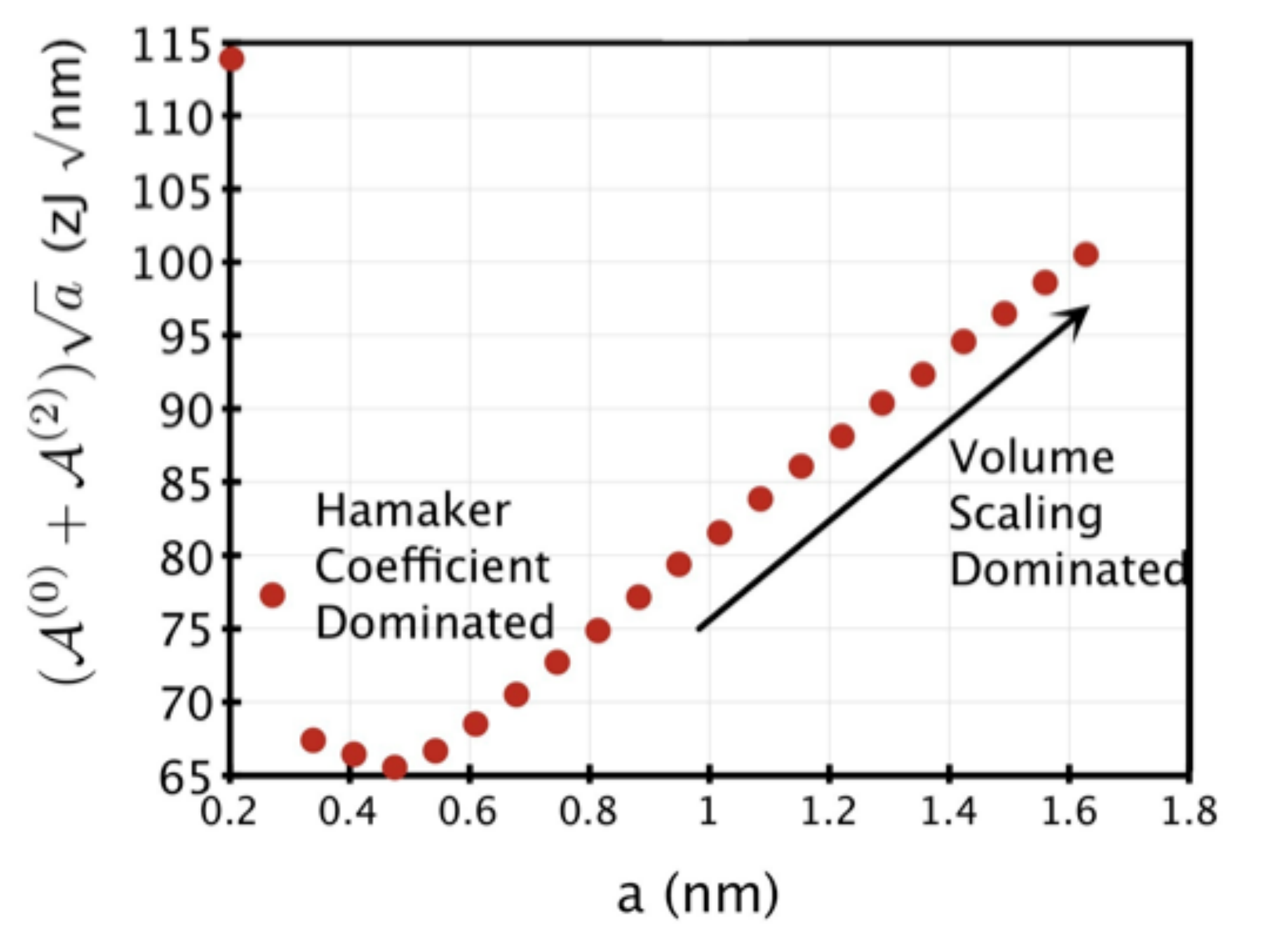} 
\caption{\RP Comparison of Hamaker coefficients multiplied by $\sqrt{a}$ for two rodlike SWCNT cylinders in the limit of small separations in water. Hamaker coefficients were calculated using the anisotropic rod-rod close separation result with armchair vdW-Ld spectra. The interaction free energy, proportional to Hamaker coefficients multiplied by $\sqrt{a}$ at a fixed spacing, distinguishes "material dominated" and "size dominated" behavior for small vs. large radii.}
\label{fig:TE_A121_Near} 
\end{figure}

Figure \ref{fig:TE_A121_Near} shows Hamaker coefficient multiplied by $\sqrt{a}$. We see two competing effects. As the Hamaker coefficients themselves drop as a function of radius and reach an asymptotic behavior around 0.8 nm radius and above, variation of the Hamaker coefficient multiplied by $\sqrt{a}$, i.e., the total interaction free energy,  for each SWCNT type results in a bowl-shaped curve. The interaction free energy at a fixed separation between two SWCNTs can thus be varied either by changing the materials, affecting the Hamaker coeffcient directly, or by changing the radius of the SWCNT. However, the SWCNT material properties and their radii are coupled. From Figure \ref{fig:TE_A121_Near} one can deduce that in the small-radius regime changes in the interaction free energy at a given surface separation are more directly correlated with the changes in the Hamaker coefficient itself, while for larger radii there is a small difference in the Hamaker coefficients but the changes in the radius can be big. We term the first type of behavior ''material dominated'' and the second one ''size dominated''.

\section{Conclusions}
\label{sec:conclusions}

We have reviewed $\varepsilon''(\omega)$  optical dispersion properties for transition energies up to 30 eV and Hamaker coefficients of the vdW - Ld interactions of 64 SWCNTs.  The resulting trends in the 0-5 eV or $\pi$ regime were found to depend on the cutting line position and packing, while the 5-30 eV or $\sigma$ regime depended largely on SWCNT geometry.  We discovered that the trends as a function of radius tend to break down and become more erratic for tubes of radius less than 0.7 nm.  The trends, breakdowns, and other interesting results can be categorized in a new SWCNT classification based on a combination of electronic structure, curvature, and geometry. Ultimately, the analysis here can lead to a further understanding of the chirality-dependent properties of SWCNTs exploitable by various experimental means. The use of {\sl ab initio} methods to obtain realistic directionally dependent optical absorption spectra for highly complex systems, that are not available by other means \cite{[63]}, can accelerate the study of vdW-Ld interactions in these systems. 

We discovered that the optical properties for SWCNTs need a new optical dispersion classification scheme in order to adequately group and address certain features that would be pertinent to practical properties. We have classified the sub-groups of SWCNTs with regard to $\sigma$, $\pi$ and metal features. These properties determine also the Hamaker coefficients and total energies of vdW-Ld interactions in many ways to serve as a basis for novel classification schemes. Standard classifications based on fewer groupings of the properties (e.g. zone-folding and Lambin classifications \cite{[39]}) or on symmetry considerations (e.g. Saito/Dresselhaus classification \cite{[15]}) would in our opinion neglect key possibilities in experimental design. Exploiting important differences in the vdW-Ld interactions of nanoscale objects will allow better separation and placement into nanodevices.

Some SWCNT classes may ultimately be redundant (e.g. the chiral zigzag semiconductors) because the differences are not significant enough to warrant such fine distinction. Our optical dispersion classification is also equally incapable of predicting the unique features that occur for the very smallest SWCNTs. For example, it cannot explain why the {[}5,0,m{]} is a metal despite having an {\RP (n,m)} vector that predicts a semiconductor. When curvature effects become large, the impact on $\varepsilon''(\omega)$ becomes too strong to allow for a generalized categorization. The optical dispersion classification also fails to explain why the large diameter zigzag metals lose their metal features. 

There are also other material properties that may require revisiting this classification system. The effects of structural relaxation \cite{[62]}, phonons \cite{[17]}, bending moments \cite{JBernholc}, functionalization \cite{SDS}, the importance of excitonic effects in the optical spectral calculation \cite{Exciton1,Exciton2,Exciton3,Exciton4,Exciton5} and interlayer coupling \cite{[48]} may in fact cause armchair tubes to lose some symmetry and regain a low-energy spike. Furthermore a high transverse electric field may in fact cause the band gaps to open up \cite{[50],Slava2,[52],[53]}. These effects are well beyond the scope of this paper, but should be kept in mind in order to see if some differentiations are redundant or if new sub-classes need to be formed. {\RP Even if changes in the metal and $\pi$ peaks do arise, the more stable $\sigma$ peaks should still justify the differentiation from at least an optical property characterization standpoint.
Features of SWCNT dielectric response justify their newly proposed dispersion classification. 
While most of this analysis was specific to SWCNTs, the general concepts described can be applied to other nano-materials where vdW-Ld spectral properties can be altered as a function of geometry or other controllable parameters.}

\section{Acknowledgments}
\label{sec:acknowledgments}

This research was supported by the U.S. Department of Energy, Office of Basic Energy Sciences, Division of Materials Sciences and Engineering under Award DE-SC0008176  and DE-SC0008068. R. Rajter acknowledges financial support for this work by the NSF Grant under Contract No. CMS-0609050 (NIRT).  This research used the resources of NERSC supported by the Office of Science of DOE under contract No. DE-AC03-76SF00098. 

The {\sl ab initio} optical properties for the SWCNTs discussed here are available as part of Gecko Hamaker, the Open Source, cross-platform program for computing van der Waals – London dispersion interaction energies and full spectral Hamaker coefficients \url{http://geckoproj.sourceforge.net/}.

\section{Table}

\begin{table}[!t]
\begin{center}
\scalebox{0.55}{%
\begin{tabular}{|c|c|c|c|c|c|c|c|c|}
\hline  
 &  &  &  & $\frac{n-m}{3}$ & & $\rm sp^2-sp^3$ & Symmetry & \\
\hline 
n & m &  radius $(\AA)$ & angle & geometry & Zone Folding & Lambin & Saito & atoms \\
\hline\hline  3 & 3 &	  2.034 &	30.00 &	armchair &	metal &	metal &	M-2p &     12  \\
\hline 4 & 4 &	  2.712 &	30.00 &	armchair & 	metal &	metal &	M-2p &	   16  \\
\hline 5 &	 5 &	  3.390 &	30.00 &	armchair & 	metal &	metal &	M-2p &	   20  \\
\hline  6 &	 6 &	  4.068 &	30.00 &	armchair & 	metal &	metal &	M-2p &	   24  \\
\hline 7 &	 7 &	  4.746 &	30.00 &	armchair & 	metal &	metal &	M-2p &	   28  \\
\hline 8 &	 8 &	  5.424 &	30.00 &	armchair & 	metal &	metal &	M-2p &	   32  \\
\hline  9 &	 9 &	  6.102 &	30.00 &	armchair & 	metal &	metal &	M-2p &	   36  \\
\hline 10 &	10 &	  6.780 &	30.00 &	armchair & 	metal &	metal &	M-2p &	   40  \\
\hline 11 &	11 &	  7.458 &	30.00 &	armchair & 	metal &	metal &	M-2p &	   44  \\
\hline 12 &	12 &	  8.136 &	30.00 &	armchair & 	metal &	metal &	M-2p &	   48  \\
\hline 13 &	13 &	  8.814 &	30.00 &	armchair & 	metal &	metal &	M-2p &	   52  \\
\hline 14 &	14 &	  9.492 &	30.00 &	armchair & 	metal &	metal &	M-2p &	   56  \\
\hline 15 &	15 &	 10.170 &	30.00 &	armchair & 	metal &	metal &	M-2p &	   60  \\
\hline 16 &	16 &	 10.848 &	30.00 &	armchair & 	metal &	metal &	M-2p &	   64  \\
\hline 17 &	17 &	 11.526 &	 30.00 &	armchair & 	metal &	metal &	M-2p &	   68  \\
\hline 18 &	18 &	 12.204 &	30.00 &	armchair & 	metal &	metal &	M-2p &	   72  \\
\hline 19 &	19 &	 12.882 &	30.00 &	armchair & 	metal &	metal &	M-2p &	   76  \\
\hline 20 &	20 &	 13.560 &	30.00 &	armchair & 	metal &	metal &	M-2p &	   80  \\
\hline 21 &	21 &	 14.238 &	30.00 &	armchair & 	metal &	metal &	M-2p &	   84  \\
\hline 22 &	22 &	 14.916 &	30.00 &	armchair & 	metal &	metal &	M-2p &	   88  \\
\hline 23 &	23 &	 15.594 &	30.00 &	armchair & 	metal &	metal &	M-2p &	   92  \\
\hline 24 &	24 &	 16.272 &	30.00 &	armchair & 	metal &	metal &	M-2p &	   96  \\
\hline  6 &	 0 &	  2.349 &	 0.00 &	  zigzag &	metal &	semimetal &	  M1	 &   24  \\
\hline  9 &	 0 &	  3.523 &	 0.00 &	  zigzag &	metal &	semimetal &	  M1	 &   36  \\
\hline 12 &	 0 &	  4.697 &	 0.00 &	  zigzag &	metal &	semimetal &	  M1 &   48  \\
\hline 15 &	 0 &	  5.872 &	 0.00 &	  zigzag &	metal &	semimetal &	  M1	 &   60  \\
\hline 18 &	 0 &	  7.046 &	 0.00 &	  zigzag &	metal &	semimetal &	  M1	 &   72  \\
\hline 21 &	 0 &	  8.220 &	 0.00 &	  zigzag &	metal &	semimetal &	  M1	 &   84  \\
\hline 24 &	 0 &	  9.395 &	 0.00 &	  zigzag &	metal &	semimetal &	  M1	 &   96  \\
\hline 27 &	 0 &	 10.569 &	 0.00 &	  zigzag &	metal &	semimetal &	  M1	 &  108  \\
\hline 30 &	 0 &	 11.743 &	 0.00 &	  zigzag &	metal &	semimetal &	  M1	 &  120  \\
\hline  7 &	 0 &	  2.740 &	 0.00 &	  zigzag &	semi &	semiconductor &	  S1 &	   28  \\
\hline  8 &	 0 &	  3.132 &	 0.00 &	  zigzag &	semi &	semiconductor &	  S2 &	   32  \\
\hline 10 &	 0 &	  3.914 &	 0.00 &	  zigzag &	semi &	semiconductor &	  S1 &	   40  \\
\hline 11 &	 0 &	  4.306 &	 0.00 &	  zigzag &	semi &	semiconductor &	  S2 &	   44  \\
\hline 13 &	 0 &	  5.089 &	 0.00 &	  zigzag &	semi &	semiconductor &	  S1 &	   52  \\
\hline 14 &	 0 &	  5.480 &	 0.00 &	  zigzag &	semi &	semiconductor &	  S2 &	   56  \\
\hline 16 &	 0 &	  6.263 &	 0.00 &	  zigzag &	semi &	semiconductor &	  S1 &	   64  \\
\hline 17 &	 0 &	  6.655 &	 0.00 &	  zigzag &	semi &	semiconductor &	  S2 &	   68  \\
\hline 19 &	 0 &	  7.437 &	 0.00 &	  zigzag &	semi &	semiconductor &	  S1 &	   76  \\
\hline 20 &	 0 &	  7.829 &	 0.00 &	  zigzag &	semi &	semiconductor &	  S2 &	   80  \\
\hline 22 &	 0 &	  8.612 &	 0.00 &	  zigzag &	semi &	semiconductor &	  S1 &	   88  \\
\hline 23 &	 0 &	  9.003 &	 0.00 &	  zigzag &	semi &	semiconductor &	  S2 &	   92  \\
\hline 25 &	 0 &	  9.786 &	 0.00 &	  zigzag &	semi &	semiconductor &	  S1 &	  100  \\
\hline 26 &	 0 &	 10.178 &	 0.00 &	  zigzag &	semi &	semiconductor &	  S2 &	  104  \\
\hline 28 &	 0 &	 10.960	& 0.00 &	  zigzag &	semi &	semiconductor &	  S1 &	  112  \\
\hline 29 &	 0 &	 11.352 &	 0.00 &	  zigzag &	semi &	semiconductor &	  S2 &	  116  \\
\hline  5 &	 2 &	  2.445 &	16.10 &	  chiral &	metal &	semimetal &	M-2m &	   52  \\
\hline  6 &	 3 &	  3.107 &	19.11 &	  chiral &	metal &	semimetal &	  M1	 &   84  \\
\hline  7 &	 4 &	  3.775 &	21.05 &	  chiral &	metal &	semimetal &	M-2p &	  124  \\
\hline  8 &	 2 &	  3.588 &	10.89 &	  chiral &	metal &	semimetal &	M-2p &	   56  \\
\hline  8 &	 5 &	  4.446 &	22.41 &	  chiral &	metal &	semimetal &	M-2m &	  172  \\
\hline  9 &	 3 &	  4.234 &	13.90 &	  chiral &	metal &	semimetal &	  M1	 &  156  \\
\hline 10 &	 4 &	  4.889 &	16.10 &	  chiral &	metal &	semimetal &	M-2m &	  104  \\
\hline 11 &	 2 &	  4.746 &	 8.21 &	  chiral &	metal &	semimetal &	M-2m &	  196  \\
\hline 11 &	 8 &	  6.468 &	24.79 &	  chiral &	metal &	semimetal &	M-2m &	  364  \\
\hline  4 &	 2 &	  2.071 &	19.11 &	  chiral &	semi &	semiconductor &	  S2 &	   56  \\
\hline  5 &	 1 &	  2.179 &	 8.95 &	  chiral &	semi &	semiconductor &	  S1 &	  124  \\
\hline  6 &	 1 &	  2.567 &	 79 &	  chiral &	semi &	semiconductor &	  S2 &	  172  \\
\hline  6 &	 2 &	  2.823 &	13.90 &	  chiral &	semi &	semiconductor &	  S1 &	  104  \\
\hline  6 &	 4 &	  3.413 &	23.41 &	  chiral &	semi &	semiconductor &	  S2 &	  152  \\
\hline  6 &	 5 &	  3.734 &	27.00 &	  chiral &	semi &	semiconductor &	  S1 &	  364  \\
\hline  7 &	 5 &	  4.087 &	24.50 &	  chiral &	semi &	semiconductor &	  S2 &	  436  \\
\hline  9 &	 1 &	  3.734 &	 5.21 &	  chiral &	semi &	semiconductor &	  S2 &	  364  \\
\hline 
\end{tabular}}
\caption{Tabulated data of all SWCNTs.  The chirality vector (n,m) determines the radius, angle, and geometry.  The classification depends on the relationship between (n,m) and symmetry. The number of atoms required for a lattice repeat in the {\sl ab initio} calculations is also included.}
\end{center}
\end{table}




\footnotesize{
\bibliographystyle{rsc} 
\bibliography{my_references} 
}

\eject
\newpage

\section{Table of contents entry}

\begin{figure*}[!t]
\includegraphics[width=10.0cm]{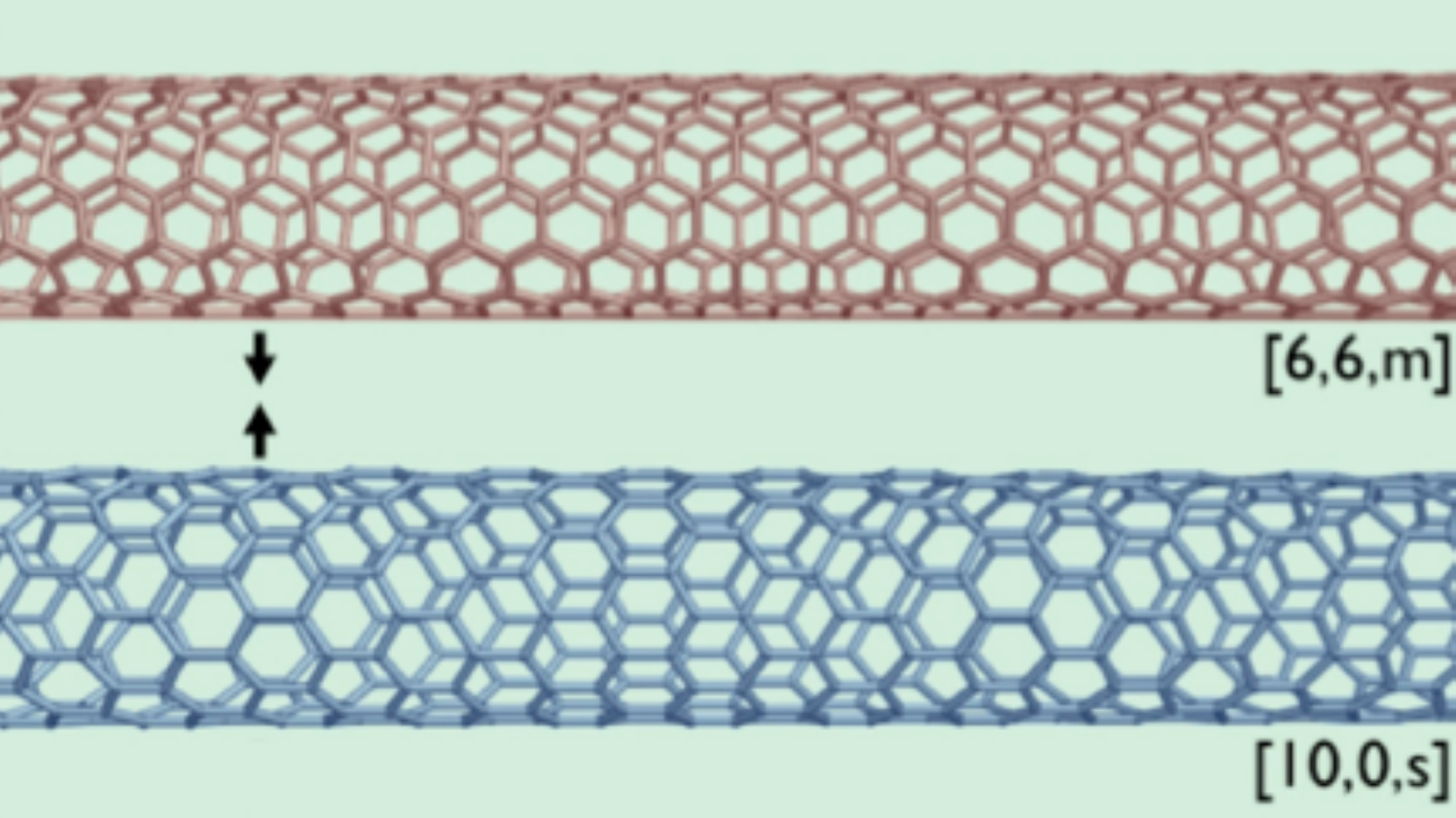} 
\caption{ {\sl Ab initio} One-electron theory based calculations to obtain the band structures  is used to obtain the frequency dependent dielectric response for 64 SWCNTs and based on this to compute their Van der Waals - London dispersion interaction energies.}
\end{figure*}


\end{document}